\documentclass[prd,onecolumn,amsmath,amssymb,floatfix,superscriptaddress,notitlepage,nofootinbib]{revtex4-1}

\usepackage{amsfonts,amssymb,amsmath,graphicx,color,bm}
\definecolor{ultramarine}{rgb}{0.07, 0.04, 0.56}
\definecolor{cadmiumgreen}{rgb}{0.0, 0.42, 0.24}
\definecolor{indigo(dye)}{rgb}{0.0, 0.25, 0.42}
\usepackage[linktocpage=true]{hyperref}
\hypersetup{
colorlinks=true,
citecolor=ultramarine,
linkcolor=cadmiumgreen,
urlcolor=indigo(dye),
}

\newcommand{\f}[2]{\frac{#1}{#2}}  
\newcommand{\mk}[1]{\left( #1 \right)}  
\newcommand{\kk}[1]{\left[ #1 \right]}  
\newcommand{\ck}[1]{\left\{ #1 \right\}}  
\newcommand{\be}{\begin{equation}}  
\newcommand{\ee}{\end{equation}}
\newcommand{\bem}{\begin{pmatrix}}
\newcommand{\eem}{\end{pmatrix}}

\renewcommand{\d}{\delta}
\newcommand{\e}{\epsilon}
\newcommand{\A}{\mathcal{A}}
\newcommand{\B}{\mathcal{B}}
\newcommand{\C}{\mathcal{C}}

\newcommand{\F}{\mathcal{F}}
\newcommand{\G}{\mathcal{G}}
\renewcommand{\L}{\mathcal{L}}

\newcommand{\R}{{R}}
\newcommand{\curv}{\mathcal{R}}
\newcommand{\mS}{\mathcal{S}}

\renewcommand{\O}{\mathcal{O}}
\newcommand{\U}{\mathcal{U}}
\newcommand{\Y}{\mathcal{Y}}
\newcommand{\Z}{\mathcal{Z}}
\newcommand{\pa}{\partial}



\begin{document}%%%%%%%%%%%%%%%%%%%%%%%%%%%%%%%%%%%%%%%%%  

\title{  
Generalized Slow Roll in the Unified Effective Field Theory of Inflation
}

\author{Hayato Motohashi}
\affiliation{Instituto de F\'{i}sica Corpuscular (IFIC), Universidad de Valencia-CSIC,
E-46980, Valencia, Spain}

\author{Wayne Hu}
\affiliation{Kavli Institute for Cosmological Physics, The University of Chicago, Chicago, Illinois 60637, USA}
\affiliation{Department of Astronomy and Astrophysics, University of Chicago, Chicago IL 60637, USA}

\begin{abstract}%%%%%%%%%%%%%%%%%%%%%%%%%%%%%%%%%%%%%%%%%  
We provide a compact and unified treatment of power spectrum observables  for the effective field theory (EFT) of inflation with the complete set of operators that lead to  second-order equations of motion
in metric perturbations in both space and time derivatives, including Horndeski and GLPV theories.
We relate the EFT operators in ADM form to the four additional free functions of time in the scalar and tensor equations.  Using the generalized slow roll formalism, we show that each power spectrum can be described by an integral over a single source that is a function of its respective sound horizon.   With this correspondence, existing model independent constraints on the source function can be simply reinterpreted in the more general inflationary context.
By expanding these sources around an optimized freeze-out epoch, we also provide  characterizations of these spectra in terms of five slow-roll hierarchies  whose leading order forms are compact and accurate as long as EFT coefficients vary only on timescales greater than an efold.
We also clarify the relationship between the unitary gauge observables employed in the EFT and the comoving gauge observables of the post-inflationary universe.
\end{abstract}

% 98.80.-k Cosmology
% 98.80.Cq Particle-theory and field-theory models of the early Universe (including cosmic pancakes, cosmic strings, chaotic phenomena, inflationary universe, etc.)  

\date{\today}

\maketitle 

%%%%%%%%%%%%%%%%%%%%%%%%%%%%%%%%%%%%%%%%%  

\section{Introduction}%%%%%%%%%%%%%%%%%%%%%%%%%%%%%%%%%%%%%%%%%
\label{sec:int}

The effective field theory (EFT) of inflation \cite{Creminelli:2006xe,Cheung:2007st} provides a general framework for understanding
the observables associated with single-field inflation.  Here a scalar field provides a clock that breaks
temporal but preserves spatial diffeomorphism invariance.   Motivated by its extension to dark energy models, subsequent
work \cite{Gleyzes:2013ooa,Kase:2014cwa,Gleyzes:2014rba,Gleyzes:2015pma}
extended the EFT to treat derivative operators that were not explicitly considered in \cite{Cheung:2007st} but arise in
Horndeski \cite{Horndeski:1974wa,Nicolis:2008in,Deffayet:2009wt,Deffayet:2009mn,Deffayet:2011gz,Kobayashi:2011nu},
Gleyzes-Langlois-Piazza-Vernizzi (GLPV) \cite{Gleyzes:2014dya,Gleyzes:2014qga} 
and Horava-Lifshitz \cite{Horava:2009uw,Blas:2009qj,Blas:2009ck} theories.

In these more general cases, the time variation of a multitude of EFT coefficients leads to a much richer range of
possibilities for the scalar and tensor power spectra, especially  beyond leading order in slow roll.  In this paper,
we undertake a unified and self-contained treatment of the general relationship between the EFT Lagrangian and
the power spectra observables.   We focus on the EFT of operators that lead to equations of motion (EOMs) for metric perturbations during inflation  that are 
second order in both time and space and hence include the Horndeski and GLPV classes.
Higher order but degenerate Lagrangians that nonetheless propagate only one extra scalar degree of freedom 
\cite{Langlois:2015cwa,Crisostomi:2016czh,Achour:2016rkg,BenAchour:2016fzp,Langlois:2017mxy}
satisfying degeneracy conditions~\cite{Motohashi:2014opa,Langlois:2015cwa,Motohashi:2016ftl,Klein:2016aiq,Crisostomi:2017aim}
and/or containing higher order spatial operators \cite{Gao:2014soa,Gao:2014fra}  are not considered here but our formalism can be
straightforwardly extended.

In \S \ref{sec:eft}, we provide a compact, self-contained and unified treatment for the quadratic Lagrangian of the EFT of inflation 
and its consequences for scalar, vector and tensor metric perturbations.      Its relationship and advantages compared
to related works \cite{Gleyzes:2013ooa,Kase:2014cwa,Gleyzes:2014rba} is explored in Appendix \ref{sec:dic}.
In \S \ref{sec:gsr}, we show that the scalar and tensor power spectra can be described in the generalized slow roll (GSR)
formalism \cite{Stewart:2001cd,Gong:2004kd,Hu:2011vr,Hu:2014hoa}
as integrals over source functions given by the EFT coefficients as long as fluctuations from scale invariance remain
small.  Existing model independent constraints on these source functions  \cite{Dvorkin:2011ui,Miranda:2014fwa} can then be simply interpreted in the general EFT, Horndeski or GLPV 
contexts.
If the EFT coefficients vary on the efold time scale or larger, these integrals can be expanded in multiple hierarchies of
slow-roll parameters.  In  \S \ref{sec:osr}, by optimizing the evaluation of these parameters, we obtain a relatively
compact but accurate description of the amplitude, tilt and running of the tilt for the scalar and tensor power
spectra in the EFT of inflation in unitary gauge.   In Appendix \ref{sec:gau}, we establish the relationship between the unitary gauge and comoving gauge curvature fluctuations which differ in the presence of EFT derivative operators.
We conclude in \S \ref{sec:con}.

Throughout the paper, 
we use the $(-+++)$ metric signature and set $M_{\rm pl} = 1/\sqrt{8\pi G}=1$.

\section{EFT of Inflation}%%%%%%%%%%%%%%%%%%%%%%%%%%%%%%%%%%%%%%%%%
\label{sec:eft}

We introduce a new notational scheme that unifies and streamlines the derivation of the quadratic action of the scalar and tensor degrees of freedom for  the EFT of inflation using its ADM form.   For the  restricted class we consider, which includes Horndeski and GLPV theories,
the resulting EOMs for metric perturbations are second order in both space and time derivatives.   Their forms are parameterized by
4 free functions of time in addition to the Hubble parameter whose evolution determines the slow-roll expansion below.   The relationship between this scheme and previous treatments in the literature~\cite{Gleyzes:2013ooa,Kase:2014cwa,Gleyzes:2014rba} is given in  Appendix~\ref{sec:dic}.

\subsection{Lagrangian}%%%%%%%%%%%%%%%%%%%%%

We begin with the $3+1$ ADM decomposition of the metric into the lapse $N$, shift $N^i$, and spatial metric
$h_{ij}$,
\be ds^2 = -N^2 dt^2 + h_{ij} (dx^i+N^idt)(dx^j+N^jdt) . 
\label{eqn:adm}
\ee
Using a unit vector $n_\mu=-Nt_{,\mu}=(-N,0,0,0)$ orthogonal to constant $t$ surfaces,
we define the acceleration $a_\mu \equiv n_{\mu;\nu} n^\nu$ and the extrinsic curvature $K_{\mu\nu}=n_{\nu;\mu}+n_\mu a_\nu$.  Semicolons on indices here and throughout denote
covariant derivatives with respect to $g_{\mu\nu}$.

In the EFT approach, we consider a general action which preserves
unbroken spatial diffeomorphisms but explicitly breaks temporal diffeomorphisms \cite{Cheung:2007st,Gubitosi:2012hu}.  
Specifically, we construct the action out of the geometric quantities of the
ADM decomposition \cite{Gleyzes:2014rba}
\be \label{LagEFT} S = \int d^4x N\sqrt{h} \, L (N, K^i_{\hphantom{i}j}, R^i_{\hphantom{i}j};t) , \ee 
where we have used $\sqrt{-g}=N\sqrt{h}$ with $h$ as the determinant of $h_{ij}$.   Purely spatial indices are
raised and lowered by $h_{ij}$.
Here $R_{ij}$ is the three-dimensional Ricci tensor and its trace $R=R^i_{\hphantom{i}i}$.
Since $h_{ij} \ne \delta_{ij}$, for notational convenience we denote throughout
\begin{equation} \label{spcon}
(T_{i \ldots j})^2 \equiv \delta^{ii'}\ldots \delta^{jj'} T_{i\ldots j} T_{i'\ldots j'} \ne
T^{i\ldots j} T_{i\ldots j}
\end{equation}
for any spatial tensor $T_{i\ldots j}$.

The Lagrangian~\eqref{LagEFT} encompasses a wide class of theories.
For example, the Einstein-Hilbert action is given by the Gauss-Codazzi relation up to 
a total derivative as
\begin{equation}
L =\frac{ {}^{(4)}R}{2} = \frac{1}{2} (K^i_{\hphantom{i}j} K^j_{\hphantom{i}i} -K^2+\R) ,
\end{equation}  where $K\equiv K^i_{\hphantom{i}i}$ and ${}^{(4)}R$ is the four-dimensional Ricci scalar.   More generally it includes models with an extra scalar degree of freedom by representing them in unitary gauge where the scalar is carried
by the metric.   
The constant $t$ surfaces are chosen to have spatially uniform scalar field $\phi=\phi(t)$
and kinetic term  $X\equiv g^{\mu\nu} \pa_\mu\phi\pa_\nu\phi = -\dot\phi^2/N^2$.
For example a minimally coupled canonical scalar field in the potential $V(\phi)$ has
\be L = \f{{}^{(4)}R}{2} - \f{X}{2} - V(\phi) 
= \f{{}^{(4)}R}{2}+ \f{\dot\phi^2(t)}{2N^2} - V(\phi(t)) .
\label{eqn:canonical}
\ee
Thus the dependence on $\phi$ and $X$ of the Lagrangian is  subsumed into the explicit time dependence and  lapse dependence of \eqref{LagEFT}.  More generally by restoring temporal diffeomorphisms with the St\"uckelberg trick or equivalently
transforming out of unitary gauge, \eqref{LagEFT} represents the scalar and tensor degrees
of freedom in Horndeski and GLPV theories (see \S \ref{ssec:noncan}).
However, the Lagrangian~\eqref{LagEFT} does not cover the spatially covariant gravity~\cite{Gao:2014soa,Gao:2014fra} as we do not allow extra spatial derivatives.  
Further, it does not include degenerate higher order scalar-tensor theories (DHOST) \cite{Langlois:2015cwa,Crisostomi:2016czh,Achour:2016rkg,BenAchour:2016fzp,Langlois:2017mxy} as their Lagrangians depend on $\dot N$.  
We leave the EFT description of these classes as future work.

To derive the quadratic action, we perturb the metric
around a spatially flat Friedmann-Lema\^itre-Robertson-Walker (FLRW) background
\be \bar N=1, \quad\bar N^i=0 ,\quad\bar h_{ij} = a^2 \delta_{ij}. \ee
The extrinsic and intrinsic curvatures of the background are given by
\be\bar K^i_{\hphantom{i}j}  = H\delta^i_{\hphantom{i}j} , \quad\bar R^i_{\hphantom{i}j}  = 0, \ee
where $H\equiv d\ln a/dt$.    Terms that are quadratic in the metric fluctuations are
at most quadratic in perturbations to the ADM variables and so it is useful to define the Taylor coefficients
evaluated on the background ``b",
\begin{align} \label{cxynot} 
L \Big|_{\rm b} &= \C, \notag\\
\f{\pa L}{\pa Y^i_{\hphantom{i}j}} \Big|_{\rm b} &= \C_Y \delta^j_{\hphantom{i}i}, \notag\\
\f{\pa^2 L}{\pa Y^i_{\hphantom{i}j} \pa Z^k_{\hphantom{i}\ell}} \Big|_{\rm b} &= \C_{YZ} \delta^j_{\hphantom{i}i} \delta^\ell_{\hphantom{i}k} + \f{\tilde\C_{YZ}}{2} (\delta^\ell_{\hphantom{\ell}i} \delta^j_{\hphantom{i}k} + \delta_{ik} \delta^{j\ell} ) ,  
\end{align}
where $Y,Z \in \{ N, K, R \}$ and the index structure is determined by the symmetry of the background.  
For notational simplicity we treat scalars and traces with the same notation; thus implicitly $N = N^i_{\hphantom{i}i}$ and $\tilde\C_{NZ} = 0$.  
Up to quadratic order
\begin{align}
L &= \C
+ \sum_{Y} \C_{Y} \delta Y 
+ \f{1}{2}\sum_{Y,Z} (\C_{YZ}\delta Y\delta Z + \tilde\C_{YZ} \delta Y^i_{\hphantom{i}j} \delta Z^j_{\hphantom{i}i} ) .
\label{eqn:Lagrangian1}
\end{align}
Note that the $\C$'s are functions of time only as they are evaluated on the background but are in general free functions in the EFT. 
In a specific model they take on definite forms, e.g.~for the Lagrangian (\ref{eqn:canonical}) of the canonical scalar field
\be \label{cano1} 
\C = -3H^2+\f{\dot\phi^2}{2} - V, \quad
\C_N = -\dot\phi^2, \quad 
\C_K = -2H, \quad 
\C_R = \f{1}{2} , \quad  
\tilde\C_{KK} = -\C_{KK}=1, \quad
\C_{NN} = 3 \dot\phi^2,
\ee
with other $\C$ functions being zero.
We provide more nontrivial examples in \S \ref{ssec:noncan}.

With these definitions we can directly evaluate the quadratic action of scalar, vector and tensor
metric perturbations.   This means that  $\delta K^i_{\hphantom{i}j}$ and  $\delta R^i_{\hphantom{i}j}$
must be in principle expanded to second order in metric fluctuations.   
Since $K^i_{\hphantom{\,}j}$ is the most complicated in terms of
metric fluctuations, it is advantageous to eliminate the linear term in $\delta K= K- 3H$ and hence the need to expand it to second order in the metric.
Since $K = n^{\mu}_{\hphantom{\mu};\mu}$, we can integrate by parts expressions of the form
\begin{equation}
\int d^4 x \sqrt{-g} F(t) K = -\int d^4 x \sqrt{-g} n^\mu F_{;\mu} = - \int d^4 x \sqrt{-g} \frac{\dot F}{N}
\end{equation}
ignoring boundary terms.
Therefore, the Lagrangian (\ref{eqn:Lagrangian1}) can be rewritten as\footnote{\label{fn1}In \cite{Gleyzes:2013ooa,Kase:2014cwa,Gleyzes:2014rba},
$\delta K^i_{\hphantom{i}j} \delta R^j_{\hphantom{i}i}$ is also integrated by parts 
and the $1/N$ term  is expanded to second order but these steps make the derivation more cumbersome; see Appendix~\ref{sec:dic} and the canceling $N\sqrt{h}$ factor in \eqref{LagEFT3}.
}
\begin{eqnarray} \label{LagEFT2}
L  =  \C  -\frac{\dot \C_K}{N} - 3 H \C_K +
\C_{N} \delta N + \C_R\delta R
+ \frac{1}{2}  \sum_{Y, Z} (\C_{YZ} \delta Y\delta Z + \tilde\C_{YZ} \delta Y^i_{\hphantom{i}j} \delta Z^j_{\hphantom{i}i}).
\label{eqn:Lagrangian2}
\end{eqnarray}
Since metric fluctuations also appear in the volume element, the quadratic action
follows from keeping terms in the quadratic terms in the expansion of ${\cal L} = N\sqrt{h}L$,
\begin{equation} 
{\cal L}
= N\sqrt{h} ( \C- 3 H \C_K) - \sqrt{h} {\dot \C_K}  +
N\sqrt{h} ( \C_{N} \delta N + \C_R\delta R) 
+ \frac{a^3}{2}  \sum_{Y, Z} (\C_{YZ} \delta Y\delta Z + \tilde\C_{YZ} \delta Y^i_{\hphantom{i}j} \delta Z^j_{\hphantom{i}i}),
\label{LagEFT3}
\end{equation}
where we have dropped terms that are manifestly higher order. 
The quadratic action can be more explicitly written by employing the background EOMs,
\begin{eqnarray}
\C - 3 H \C_K + \C_N &=& 0 ,\nonumber\\
\C - 3 H \C_K - \dot \C_K &=& 0,
\label{eqn:background}
\end{eqnarray}
which come from the first order variation with respect to the ADM variables that are allowed by the symmetries of the background, 
$N$ and $\sqrt{h} = a^3$.  The background equations imply $\C_N = -\dot \C_K$ for the EFT of inflation where there
are no other matter species (but not for the EFT of dark energy \cite{Gleyzes:2013ooa}).  
Note that the term linear in $\delta R$ is a total
spatial derivative term on the background that does not produce an extra background EOM.
For example in the canonical scalar case~\eqref{cano1}, the background equations 
(\ref{eqn:background}) are given by 
\begin{align}
3H^2 &= \f{\dot\phi^2}{2} + V, \notag\\
3H^2 + 2\dot H &= -\f{\dot\phi^2}{2} + V ,
\end{align}
as expected. 
Employing the background EOMs in the Lagrangian \eqref{LagEFT3},
we obtain a relatively compact and transparent form for the quadratic action 
\be
a^{-3} {\cal L}_2 = \C_N (\delta N)^2 + \C_R \kk{ \mk{ \delta N + \frac{\delta \sqrt{h}}{a^3} } \delta_1 R + \delta_2 R }
+ \frac{1}{2}  \sum_{Y, Z} (\C_{YZ} \delta Y\delta Z + \tilde\C_{YZ} \delta Y^i_{\hphantom{i}j} \delta Z^j_{\hphantom{i}i}) .
\label{LagEFT4}
\ee
Note that $\delta R= \delta_1 R + \delta_2 R + \ldots$ where the terms denote the contributions that
are  first and second order in the underlying scalar, vector and tensor
metric perturbations that we consider next.

\subsection{Tensor perturbation}%%%%%%%%%%%%%%%%%%%%%

First, we consider the tensor perturbation in the ADM metric 
\be N = 1, \quad N_i = 0 , \quad  h_{ij} = a^2(\delta_{ij}+\gamma_{ij}) , \ee
where the spatial metric fluctuation is transverse-traceless $\delta^{ij}\gamma_{ij} = \delta^{ij}\pa_{i}\gamma_{jk}=0$.   The ADM curvature perturbations then become
\begin{align}
\delta  K^i_{\hphantom{i}j} &= \f{1}{2} \dot \gamma^i_{\hphantom{i}j}, \notag\\
\delta_1 R^i_{\hphantom{i}j} &= 0, \notag\\
\delta_2 R &= \f{1}{a^2} \delta^{ii'}\delta^{jj'}\delta^{kk'} \mk{ \gamma_{ij} \pa_k\pa_{k'} \gamma_{i'j'} + \f{3}{4}  \pa_k \gamma_{ij} \pa_{k'} \gamma_{i'j'} - \f{1}{2} \pa_k \gamma_{ij} \pa_{j'} \gamma_{i'k'} } \sim - \f{1}{4a^2} (\pa_k \gamma_{ij})^2,
\end{align}
where 
we used integration by parts in the last equality which holds even in the presence of a prefactor depending on $t$ and recall the notation~\eqref{spcon} for the contraction of a squared tensor.
The quadratic Lagrangian (\ref{LagEFT4}) becomes
\be \label{S2t-1}
{\cal L}_2
= a^3 \left[ \frac{\tilde\C_{KK}}{8} \dot\gamma_{ij}^2 - \frac{\C_R }{4a^2}( \pa_k \gamma_{ij} )^2  \right] .
\ee
We can further simplify the Lagrangian in terms of the amplitude of the two gravitational wave polarization states
of  wavenumber $k$
\be 
\label{S2t}
{\cal L}_2  = \sum_{\lambda=+,\times} \f{a^3 b_t}{4c_t^2} \mk{ \dot\gamma_\lambda^2 - \f{c_t^2k^2}{a^2} \gamma_\lambda^2} , \ee
where 
\be b_t = 2\C_R , \quad c_t^2 = \f{2\C_R}{\tilde\C_{KK}} . \ee
For example for a gravitational wave traveling in the $z$ direction
\be
\gamma_{ij}(t,z) = \gamma_+(t) e^{ikz}(\delta_{i x}\delta_{j x} - \delta_{i y} \delta_{j y})
+\gamma_\times(t) e^{i kz} (\delta_{i x}\delta_{j y} + \delta_{j x} \delta_{i y}).
\ee 
Evidently, $c_t$ plays the role of the sound speed for tensor perturbations.  
We have written the normalization factor as $b_t$ rather than using $\tilde \C_{KK}$ so as to parallel our treatment of scalars below. 
Note that $b_t=c_t=1$ for the canonical case~\eqref{cano1} and so their time dependence in the
EFT of inflation leads to new slow roll hierarchies.

\subsection{Vector perturbation}%%%%%%%%%%%%%%%%%%%%%

We can use gauge freedom to remove the vector perturbation to
the three-dimensional metric $h_{ij}$ leaving the ADM metric 
\be N=1, \quad N_i = v_i , \quad h_{ij} = a^2 \delta_{ij},  \ee
with $\delta^{ij} \pa_i v_j=0$.  
Imposing this gauge fixing at the action level does not lose any independent EOMs~\cite{Motohashi:2016prk}.
Since
\be \delta K_{ij} = -\f{1}{2} (\pa_i v_j + \pa_j v_i), \ee
the quadratic Lagrangian is given by 
\be \L_2 = \f{ \tilde\C_{KK}}{8a }\left({\pa_iv_j + \pa_jv_i}\right)^2. \ee
Vector perturbations are non-dynamical and with no source in the matter sector can consistently be set to zero.

\subsection{Scalar perturbations}%%%%%%%%%%%%%%%%%%%%%

For the scalar perturbations, the assumption of unitary gauge in the EFT Lagrangian \eqref{LagEFT} fixes the temporal gauge freedom.   
To fully remove the gauge freedom, and allow the gauge to be fixed at the action level~\cite{Motohashi:2016prk}, we take the ADM metric to be given by
\be \label{unimet} N = 1+\delta N, \quad N_i = \pa_i \psi, \quad h_{ij} = a^2e^{2\zeta} \delta_{ij} .\ee
We discuss its
relationship to alternate gauges, especially the comoving gauge, in 
Appendix~\ref{sec:gau}.

The ADM volume and curvature perturbations are then
\begin{align}
\delta \sqrt{h} &= 3a^3 \zeta, \notag\\
\delta  K^i_{\hphantom{i}j} &= (\dot\zeta-H\delta N)\delta^i_{\hphantom{i}j} -\f{1}{a^2}\delta^{ik}\pa_k\pa_j \psi, \notag\\
\delta  K &= 3(\dot\zeta-H\delta N) - \f{\pa^2 \psi}{a^2}, \notag\\
\delta_1 R^i_{\hphantom{i}j} &= - \f{1}{a^2} ( \delta^i_{\hphantom{i}j} \pa^2 \zeta + \delta^{ik} \pa_k \pa_j \zeta) ,\notag\\ 
\delta_2 R &= -\f{2}{a^2} [(\pa\zeta)^2-4\zeta \pa^2\zeta ] 
\sim -\f{10}{a^2} (\pa\zeta)^2,
\end{align}
where the notation~\eqref{spcon} implies $\pa^2 = \delta^{ij} \partial_i\partial_j$ and $(\pa\zeta)^2 = \delta^{ij} \pa_i\zeta \pa_j\zeta$.
Note that through integration by parts
\begin{align}
\delta K^i_{\hphantom{i}j} \delta   K^j_{\hphantom{i}i} &\sim 3(\dot\zeta-H\delta N)^2 - 2(\dot\zeta-H\delta N) \f{\pa^2\psi}{a^2} + \mk{\f{\pa^2\psi}{a^2}}^2, \notag\\
\delta  K^i_{\hphantom{i}j} \delta_1 R^j_{\hphantom{i}i} &\sim - 4(\dot\zeta-H\delta N) \f{\pa^2\zeta}{a^2} + 2 \f{\pa^2\psi}{a^2} \f{\pa^2\zeta}{a^2}  , \notag\\
\delta_1 R^i_{\hphantom{i}j} \delta_1 R^j_{\hphantom{i}i} &\sim 6 \mk{ \f{\pa^2 \zeta}{a^2} }^2  .
\end{align}
The quadratic Lagrangian~\eqref{LagEFT4} thus reads 
\begin{align} 
\L_2 
&= a^3 \left[ 
\left( \f{1}{2} \C_{NN}+ \C_N \right) \delta N^2 
+ \ck{ \C_{NK} \left[ 3(\dot\zeta-H\delta N) - \f{\pa^2 \psi}{a^2} \right] - 4(\C_{NR} + \C_R) \f{\pa^2 \zeta}{a^2}  } \delta N 
+ 2 \C_R \f{(\pa\zeta)^2}{a^2}  \right. \notag\\
&\qquad 
+ \f{3}{2}(3\C_{KK}+\tilde\C_{KK})(\dot\zeta-H\delta N)^2 - (3\C_{KK}+\tilde\C_{KK})(\dot\zeta-H\delta N) \f{\pa^2\psi}{a^2} + \f{1}{2}(\C_{KK}+\tilde\C_{KK})\mk{\f{\pa^2\psi}{a^2}}^2 
\notag\\
&\qquad \left. 
- 4 (3\C_{KR}+\tilde\C_{KR}) (\dot\zeta-H\delta N) \f{\pa^2\zeta}{a^2} + 2 (2\C_{KR}+\tilde\C_{KR}) \f{\pa^2\psi}{a^2} \f{\pa^2\zeta}{a^2} 
+ (8\C_{RR} + 3\tilde\C_{RR}) \mk{ \f{\pa^2 \zeta}{a^2} }^2 \right]  .
\end{align}

In the analysis below, we restrict our consideration to theories with no more than second order spatial derivatives in the EOMs of perturbations which include the Horndeski and GLPV classes.    
In this case the Lagrangian satisfies the following conditions \cite{Gleyzes:2013ooa}
\be \label{eqn:conditions} \tilde\C_{KK} = -\C_{KK} ,\quad 
\tilde\C_{KR} = -2\C_{KR} ,\quad 
\tilde\C_{RR} = -\f{8}{3}\C_{RR} . \ee
Under this set of assumptions \eqref{eqn:conditions}, 
the scalar quadratic Lagrangian becomes
\begin{align} 
\L_2 
&= a^3 \left[ 
\left( \f{1}{2} \C_{NN}+ \C_N \right) \delta N^2 
+ \ck{ \C_{NK} \mk{ 3(\dot\zeta-H\delta N) - \f{\pa^2 \psi}{a^2} } - 4(\C_{NR} + \C_R) \f{\pa^2 \zeta}{a^2}  } \delta N 
+ 2 \C_R \f{(\pa\zeta)^2}{a^2} \right. \notag\\
&\qquad \left. 
+ 3\C_{KK} (\dot\zeta-H\delta N)^2 - 2\C_{KK}(\dot\zeta-H\delta N) \f{\pa^2\psi}{a^2} 
- 4 \C_{KR} (\dot\zeta-H\delta N) \f{\pa^2\zeta}{a^2} \right] .
\end{align}
Furthermore the Hamiltonian and momentum constraints render the lapse 
and shift to be non-dynamical as usual. 
Indeed the EOMs for $\psi$ and $\delta N$ are given by
\begin{align} 
\label{coneq-lapse} \delta N &= \f{2\C_{KK}}{2H\C_{KK}-\C_{NK}} \dot\zeta, \notag\\
\f{\pa^2\psi}{a^2} &= - \f{1}{2H\C_{KK} - \C_{NK}} \kk{ (\C_{NN}+2 \C_N) \delta N - 3(2H\C_{KK}-\C_{NK}) (\dot\zeta - H \delta N) + 4 (H\C_{KR} - \C_{NR} - \C_{R}) \f{\pa^2\zeta}{a^2} } . 
\end{align}
We therefore also assume
\be \label{dencond} 2H\C_{KK}-\C_{NK} \ne 0 . \ee
Given that this condition involves $H$, it is a property of the background solution and cannot be imposed directly on a scalar field Lagrangian  in contrast to (\ref{eqn:conditions}).
As shown in Appendix~\ref{sec:gau},  violation of (\ref{dencond}) is associated with 
unitary gauge being ill-defined [see \eqref{Gam}], which indicates that constant field slices are no longer spacelike Cauchy surfaces.  We thus assume the condition (\ref{dencond}) is satisfied for the following analysis.

Eliminating the lapse and shift brings the quadratic Lagrangian of the remaining variable $\zeta$ to
\begin{align} \label{S2s-1} 
\L_2 
&= a^3 \kk{ \A_{\dot\zeta \dot\zeta } \dot\zeta^2 
- 2 \A_{\dot\zeta \zeta} \f{\dot\zeta \pa^2 \zeta }{a^2}  
+ \A_{\zeta\zeta} \f{(\pa \zeta)^2}{a^2}   },
\end{align} 
where
\begin{eqnarray}
{\cal A}_{\dot\zeta\dot\zeta} &=& \frac{ \C_{KK} [ 2 \C_{KK} (\C_{NN}+2 \C_{N} )- 3 \C_{NK}^2 ] }{(2 H \C_{KK}-\C_{NK})^2} , \nonumber\\
{\cal A}_{\dot\zeta\zeta} &=&   \frac{ 4 \C_{KK} (\C_R + \C_{NR}) - 2 \C_{KR} \C_{NK} }{ 2 H \C_{KK} - \C_{NK} } , \nonumber\\
{\cal A}_{\zeta\zeta} &=& 2  \C_{R} .
\end{eqnarray}
Using integration by parts,
the quadratic action in Fourier space is given by 
\be \label{S2s} S_2 = \int d^4x~ \f{a^3 b_s \e_H}{c_s^2} \mk{ \dot\zeta^2 - \f{c_s^2k^2}{a^2}\zeta^2}, \ee
where $\e_H= -\dot H/H^2$,  
\begin{align} 
b_s &\equiv -\f{1}{\epsilon_H}\mk{ \A_{\zeta\zeta} - H \A_{\dot\zeta\zeta} - \dot \A_{\dot\zeta\zeta} } ,\notag\\
c_s^2 &\equiv - \A_{\dot\zeta \dot\zeta }^{-1} \mk{ \A_{\zeta\zeta} - H \A_{\dot\zeta\zeta} - \dot \A_{\dot\zeta\zeta} } . 
\end{align}
Note that the relation 
\be \label{bscsrel} b_s = \f{\A_{\dot \zeta\dot\zeta}}{\e_H} c_s^2 , \ee
 holds by definition. 
Evidently, $c_s$ plays the role of the sound speed for scalar perturbations.
For the canonical case \eqref{cano1}, $b_s=c_s=1$.
In the notation of
\cite{Kase:2014cwa}, the term in the prefactor of the quadratic action is used 
directly  $Q_s= \A_{\dot \zeta\dot\zeta}$.    We choose to separate these contributions to highlight deviations from the
canonical case and their role in the   slow roll expansion.

\subsection{Non-canonical examples}%%%%%%%%%%%%%%%%%%%%%
\label{ssec:noncan}

In the canonical case  \eqref{cano1}, $b_s=c_s=b_t=c_t=1$, and so the only slow-roll function upon which to develop a
slow-roll hierarchy during inflation is
the Hubble parameter $H$ itself.   More generally each of these functions is endowed with a slow-roll hierarchy of its own as we shall see below.  Although we are mainly interested in a model
independent description of inflationary observables, it is useful first
to consider examples of model classes that provide non-trivial values for these 4 free functions.

For a $P(X,\phi)$ model where recall $X=-\dot\phi^2/N^2$,
\be L = \f{{}^{(4)}R}{2}  + P(X,\phi), \ee
and we have 
\begin{align} \label{PXp} 
&\C = -3H^2+P, \quad
\C_N = -2 X P_{,X} , \quad 
\C_K = -2H, \quad 
\C_R = \f{1}{2} , \quad 
\tilde\C_{KK} = -\C_{KK}=1, \quad
\C_{NN} = 4 X^2 P_{,XX}+ 6 X P_{,X},
\end{align}
with other functions being zero,
which implies 
\be
{\cal A}_{\dot\zeta\dot\zeta} =  \frac{\C_{NN}+ 2 \C_N}{ 2 H^2} ,\quad
{\cal A}_{\dot\zeta\zeta} = H^{-1} ,\quad
{\cal A}_{\zeta\zeta} = 1 ,
\ee
and
\begin{equation}
c_s^2= \frac{2 H^2}{\C_{NN} + 2 \C_N} \frac{ d H^{-1}}{d t} = \frac{2H^2  }{\C_{NN} + 2 \C_N}\epsilon_H .
\end{equation}
We can further simplify the sound speed for $P(X,\phi)$ by noting that the
background equations~\eqref{eqn:background} imply $\C_N= -2\epsilon_H H^2$, 
\begin{equation}
c_s^2 = \frac{P_{,X}}{ 2 X P_{,XX} + P_{,X}},
\end{equation}
which is the expected result.  
Furthermore, from \eqref{bscsrel} we obtain
$b_s=1$ and since $P(X,\phi)$ does not contain $K$ or $R$ dependence
$b_t=c_t=1$.

In order to change $b_s$, $b_t$ and $c_t$ we need more complicated couplings in
the EFT Lagrangian involving $K$ and $R$.   A simple example is 
\be \label{Lf3} L = \f{{}^{(4)}R}{2}  + f_3\frac{K}{N^2}, \ee
where $f_3=\,$const.
In this case the non-vanishing coefficients are
\begin{align} \label{f3} 
&\C = -3H^2+3 f_3 H, \quad
\C_N = -6 f_3 H , \quad 
\C_K = -2H + f_3, \quad 
\C_R = \f{1}{2} ,  \notag\\
&\tilde\C_{KK} = -\C_{KK}=1, \quad
\C_{NN} = 18 f_3 H, \quad
\C_{NK} = -2 f_3 . 
\end{align}
Because of the nonvanishing $\C_{NK}$ term, $b_s \ne 1$ in addition to $c_s \ne 1$,
whereas $b_t=c_t=1$.

The tensor structure can be changed by altering the intrinsic curvature terms, for
example
\be \label{Lf4} 
L = \f{{}^{(4)}R}{2}  + f_4\frac{R}{N^2}, \ee
with $f_4=\,$const., where the non-vanishing coefficients are
\begin{align} \label{f3} 
&\C = -3H^2, \quad
\C_K = -2H , \quad 
\C_R = \f{1}{2} + f_4 ,  \quad
\tilde\C_{KK} = -\C_{KK}=1 , \quad  
\C_{NR} = -2 f_4.
\end{align}
Here the change in $\C_R$ allows $b_t\ne 1$ and $c_t\ne 1$ in addition to $b_s \ne 1$
and $c_s \ne 1$ due to $\C_{NR}$
(see \cite{Alishahiha:2011yh,Alishahiha:2013nsa} for a similar model motivated by asymmetric scalings in time and space in a higher dimensional theory).

These more complicated cases are members of scalar-tensor theories from the GLPV class \cite{Gleyzes:2014dya}
\begin{align} 
\label{GLPV}
L &= G_2 + G_3 \Box \phi 
+ G_4 {}^{(4)}R - 2 G_{4,X} [ (\Box\phi)^2 -\phi^{;\mu\nu}\phi_{;\mu\nu}  ] 
+ F_4 \epsilon^{\mu\nu\rho}_{\hphantom{\mu\nu\rho}\sigma} \epsilon^{\tilde\mu\tilde\nu\tilde\rho\sigma} \phi_{;\mu}\phi_{;\tilde\mu}
\phi_{;\nu\tilde\nu} \phi_{;\rho\tilde\rho}
\notag\\
&~~~~+ G_5 {}^{(4)} G^{\mu\nu} \phi_{;\mu\nu} + \f{1}{3}G_{5,X} [ (\Box\phi)^3 - 3 \Box\phi \phi_{;\mu\nu}\phi^{;\mu\nu} + 2 \phi_{;\mu\nu}\phi^{;\mu\sigma} \phi^{;\nu}_{\hphantom{\nu};\sigma}]  \notag\\
&~~~~+ F_5 \epsilon^{\mu\nu\rho\sigma }\epsilon^{\tilde \mu\tilde\nu\tilde\rho\tilde \sigma} 
\phi_{;\mu}\phi_{;\tilde \mu} \phi_{;\nu\tilde\nu} \phi_{;\rho\tilde\rho} \phi_{;\sigma\tilde\sigma},
\end{align} 
where the $G_i$ and $F_i$ are general functions of $\phi, X$ and $\epsilon^{\mu\nu\rho\sigma}$ is the totally
antisymmetric tensor. 
In ADM form this class has the Lagrangian 
\begin{align} 
L &= A_2(t,N) + A_3(t,N)K 
+ A_4(t,N) (K^2 - K^i_{\hphantom{i}j} K^j_{\hphantom{i}i}) + B_4 (t,N) R \notag\\
&~~~~+ A_5(t,N) ( K^3 - 3K K^i_{\hphantom{i}j} K^j_{\hphantom{i}i} + 2 K^i_{\hphantom{i}j} K^j_{\hphantom{i}k} K^k_{\hphantom{i}i} ) + B_5(t,N) ( K^i_{\hphantom{i}j}R^j_{\hphantom{i}i}-\tfrac{1}{2}KR ),
\end{align}
where \cite{Gleyzes:2014dya} 
\begin{align}
A_2 &= G_2 -\sqrt{-X} \int dX \f{G_{3,\phi}}{2\sqrt{-X}}, \notag\\
A_3 &= -\int dX \sqrt{-X} G_{3,X} - 2 \sqrt{-X} G_{4,\phi} , \notag\\
A_4 &=-G_4 + 2 X G_{4,X} +\frac{X}{2}G_{5,\phi} - X^2 F_4, \notag\\
A_5 &= -\frac{1}{3} (-X)^{3/2} G_{5,X} + (-X)^{5/2} F_5 ,\notag\\
B_4 &= G_4 +\sqrt{-X} \int dX \f{G_{5,\phi}}{4\sqrt{-X}}, \notag\\
B_5 &= -\int dX \sqrt{-X} G_{5,X} . 
\end{align}
We can  see that the canonical and $P(X,\phi)$  models are represented by $G_2$ or $A_2$ and the
models of \eqref{Lf3} and \eqref{Lf4} can be described by the $A_3$ and $A_4, B_4$ or equivalently
the $G_3$ and $G_4, F_4$ functions respectively. 
It is also now clear that the EFT Lagrangian \eqref{LagEFT} can represent the whole GLPV class along with its Horndeski subset where $F_4=F_5=0$.

\section{Integral Solutions for EFT Power Spectra}%%%%%%%%%%%%%%%%%%%%%%%%%%%%%%%%%%%%%%%%%
\label{sec:gsr}

In this section, we give the scalar and tensor power spectra that result from
their respective  quadratic actions \eqref{S2s} and \eqref{S2t}.  We leave $b_s,c_s,b_t,c_t,H$ as free functions of time
in the EFT so as to keep our discussion model-independent.  We show that for small but not necessarily slowly varying deviations from scale invariance
each power spectrum is given by 
a temporal integral over a single source function formed out of a combination of these quantities.

\subsection{Scalar perturbation}%%%%%%%%%%%%%%%%%%%%%

Let us reexpress the curvature perturbation in the general
quadratic action  for scalar perturbations \eqref{S2s} by 
defining the canonically normalized scalar $u=z\zeta$ and $z=a\sqrt{2b_s \e_H}/c_s$.
We then obtain the standard Mukhanov-Sasaki equation for noncanonical inflation
\be \label{MSeqEFT} \f{d^2u}{d\eta^2} + \mk{ c_s^2 k^2 - \f{1}{z}\f{d^2z}{d\eta^2} } u = 0 , \ee
where $\eta$ is the (positive, decreasing) conformal time to the end of inflation $\eta = \int_t^{t_{\rm end}} dt/a$.
First, note that above the sound horizon $x=ks_s \ll 1$, where
\be s_s \equiv \int c_s d\eta = \int^{a_{\rm end}}_a \f{da}{a} \f{c_s}{aH} , \ee
the modefunction $u$ leaves the oscillatory regime and enters into a regime where
\begin{equation}
\frac{u}{z}  \approx c_1 + c_2 \int \frac{d\eta}{z^2},
\label{eqn:superhorizonu}
\end{equation}
or
\be \label{shsol-s} \zeta \approx c_1 + c_2 \int dt \f{c_s^2}{a^3 b_s \e_H} ,  \ee
where $c_1$ and $c_2$ are constants.

Usually we expect that the second mode is decaying on superhorizon scales and if so 
\eqref{eqn:superhorizonu} implies that the curvature perturbation $\zeta=$\,const. above the sound horizon.
However, this is not necessarily the case, even within the canonical inflation case if
the potential is exactly constant, dubbed ultra-slow-roll inflation~\cite{Kinney:2005vj}.  In  ultra-slow-roll inflation $b_s=c_s=1$ and $\e_H\propto a^{-6}$, which leads to the second mode of \eqref{shsol-s} growing.  In this case,  the consistency relation between the power spectrum and bispectrum is violated as is the separate universe condition upon which it is based~\cite{Namjoo:2012aa}.  More generally, so-called constant-roll condition $\ddot\phi=\beta H\dot\phi$ leads to $\e_H\propto a^{2\beta}$.  Therefore, if canonical inflation approaches a de Sitter expansion with $\beta<-3/2$, the curvature perturbation possesses the growing mode on superhorizon scales~\cite{Martin:2012pe,Motohashi:2014ppa,Motohashi:2017aob}.
In the more general Horndeski and GLPV classes, there are other ways in which the curvature
perturbation can grow outside the sound horizon involving $b_s$ (see \cite{Motohashi:2017vdc} for the constant-roll model in $f(R)$ gravity) but we hereafter restrict our consideration to
cases where it does not.

We can then solve \eqref{MSeqEFT} in a generalized slow roll expansion by rewriting it as
\be \label{MSeqGSR} \f{d^2y}{dx^2} + \mk{1-\f{2}{x^2}} y = \f{f''-3f'}{f} \f{y}{x^2} \ee
with
\be y \equiv \sqrt{2c_sk} ~ u, \quad
f\equiv 2\pi z \sqrt{c_s} s_s= \sqrt{ 8\pi^2 \f{b_s\e_Hc_s}{H^2} } \f{aHs_s}{c_s}.
\label{fs} \ee
Here and below $'=d/d\ln x$ but note that $x=k s_s$ and so for a given mode, the corresponding
epoch during inflation differs between scalars and tensors due to their different sound speeds.

If the curvature perturbation is frozen outside of the sound horizon,
its power spectrum reaches a well-defined limit  
\be \Delta_\zeta^2 = \lim_{x\to 0} \f{k^3}{2\pi^2} |\zeta|^2 = \lim_{x\to 0} \left| \f{xy}{f} \right|^2 , \ee
which is a natural generalization of Eq.~(22) in \cite{Hu:2011vr}.     We comment on the
relationship between the unitary gauge curvature power spectrum and the comoving
gauge curvature power spectrum that is usually taken to be the initial conditions for 
predicting scalar observables in Appendix \ref{sec:gau}.

Eq.~\eqref{fs} is exact in linear theory but not given in closed form.
However if the right hand side of \eqref{MSeqGSR} is a small source of modefunction excitations from the
Bunch-Davies vacuum form
\be \label{BDvacGSR} y_0 = \mk{1+\f{i}{x}} e^{ix} \ee
then the modefunction can be solved perturbatively.
Note that to the lowest
order in the excitations and if $f$ and the functions on which it depends are  nearly
constant 
\be
\label{eqn:curvatureleading}
\Delta_\zeta^2 \approx \frac{1}{f^2} \approx \frac{H^2}{8\pi^2 b_s\epsilon_H c_s},
\ee
which is the result given in   \cite{Kase:2014cwa}.  We separate these two pieces into the
approximation below and relax the assumptions on the constancy of the source.

\subsection{Tensor perturbations}%%%%%%%%%%%%%%%%%%%%%

The same considerations apply to tensor modes governed by \eqref{S2t} with the
canonically normalized field  $u=z\gamma_{+,\times}$, 
\begin{align} 
z&\equiv \frac{a}{c_t}\sqrt{\frac{b_t}{2}}\notag\\ 
x&\equiv ks_t = k \int dt \f{c_t}{a}, \notag\\
y&\equiv \sqrt{2c_tk} ~ u , \notag\\
f&\equiv 2 \pi z \sqrt{c_t} s_t = \sqrt{ 2\pi^2 \f{b_tc_t}{H^2} } \f{aHs_t}{c_t} .
\label{ft}
\end{align} 
which brings the EOM and the Bunch-Davies vacuum to the standard form \eqref{MSeqGSR} and \eqref{BDvacGSR} 
and generalizes \cite{Gong:2004kd,Hu:2014hoa}.

Above their sound horizon $k s_t \ll 1$, solutions take  the same form as
given by \eqref{eqn:superhorizonu} or 
\be \label{shsol-t} \gamma_{+,\times} \approx c_1 + c_2 \int dt \f{c_t^2}{a^3 b_t} . \ee
For canonical inflation $b_t=c_t=1$ and so the second term always decays with the expansion.
In principle in the Horndeski and GLPV theories
it is possible to have
tensors grow outside their sound horizon while the scalars are frozen.

Assuming that the second mode decays above the sound horizon, we reach a well-defined limit
for the tensor power spectrum sufficiently after sound horizon crossing
\be \Delta_{\gamma}^2 = \lim_{x\to 0} \f{k^3}{2\pi^2} |\gamma_{+,\times}|^2 = \lim_{x\to 0} \left| \f{xy}{f} \right| ^2 . \ee
To the lowest order in slow roll
\be \Delta_{\gamma}^2 \approx \frac{1}{f^2} \approx \frac{H^2}{2\pi^2 b_t c_t}, \ee
which again recovers the standard result.    We now generalize these tensor and scalar results for the case where the 
slow-roll functions $H,b_t, c_t, b_s, c_s$ vary with time.

\subsection{Generalized slow roll}%%%%%%%%%%%%%%%%%%%%%
\label{sec:GSR}

For both scalar and tensor perturbations, the respective power spectra $\Delta^2$
can be evaluated by solving the evolution equation \eqref{MSeqGSR} out to $x \ll 1$ with the boundary condition \eqref{BDvacGSR} at $x\to\infty$.  Beyond the leading order slow roll
approximations, these solutions can be characterized by an expansion in the observationally small deviations
from scale invariance.
One can implement this expansion systematically with the Green function technique by regarding the $f$ term
as a source of modefunction excitations away from $y_0$.

The exact, but formal, solution to \eqref{MSeqGSR} is given by \cite{Stewart:2001cd}
\be y(x) = y_0(x) - \int^\infty_x \f{dw}{w^2} \f{f''-3f'}{f} y(w) {\rm Im} [y_0^*(w)y_0(x)] . 
\label{eqn:gsriteration}
\ee
If the deviations of $y$ from $y_0$ are small, then we can replace $y \rightarrow y_0$ on the
right hand side and iteratively improve the solution.  
The first order iteration yields \cite{Kadota:2005hv}
\be \label{intform} 
\ln\Delta^2 \approx - \int^\infty_0 \f{dx}{x} W'(x) G(\ln x) ,   
\ee
where $W$ is a window function that determines the freezeout of the excitations
\be W \equiv \f{3\sin 2x}{2x^3} - \f{3\cos 2x}{x^2} - \f{3\sin 2x}{2x} ,  \ee
from the source function 
\be G \equiv -2\ln f + \f{2}{3} (\ln f)' . \ee
Since $W(0)=1$, if $f =$\,const.\ then $\Delta^2 = 1/f^2$ as expected.  
Note that   the 5 free functions $H, c_s, b_s, c_t, b_t$ are encoded into the two source functions for the power spectrum observables,
$G_\zeta(\ln x)$ for the curvature perturbation and $G_\gamma(\ln x)$ for the two tensor polarization states.

The GSR integral formula \eqref{intform} thus generalizes the slow roll approximation by 
only assuming the excitations are small in amplitude
rather than additionally assuming that their 
sources are constant or slowly varying. 
We can take the amplitude to be of order ${\cal O}(1/N)$, where the efolds are measured to the
end of inflation.  This assumption is consistent with fluctuations on the
scales observable in the CMB and large scale structure
where $N \sim 60$.   The sources, on the other hand can vary on a shorter efolding scale $\Delta N$.  
In the rest of this work, we shall consider the case where $1 \lesssim \Delta N \le N$.     We shall
see that in this case, one can Taylor expand the source in the integral.  This creates  a hierarchy of terms
separated by $1/\Delta N$ rather than $1/N$ as is assumed in the ordinary slow roll approximation.
For rapid variation $\Delta N < 1$, the opposite approximation applies since the source is more rapidly varying than the window function
\cite{Adshead:2011jq,Miranda:2012rm,Miranda:2015cea}.
For $\Delta N \sim 1$, numerical integration of \eqref{intform} is generally required.  Using our formulation,
model independent constraints from the CMB on the time variation of the
scalar source function using principal components can be simply reinterpreted in the EFT, Horndeski or GLPV context without 
requiring reanalysis of the data \cite{Dvorkin:2011ui,Miranda:2014fwa}.

\section{Optimized Slow-Roll Hierarchy for EFT}%%%%%%%%%%%%%%%%%%%%%%%%%%%%%%%%%%%%%%%%%
\label{sec:osr}

For the case in which all of the temporal variations in the source functions for the scalar
and tensor power spectra occur on the efolding scale or longer  $\Delta N>1$, the GSR integral
expression (\ref{intform}) can be analytically approximated from the Taylor expansion  of
the sources around the freezeout epoch forming a hierarchy of slow roll parameters.
The separation in amplitude between terms in this 
hierarchy is $1/\Delta N$ and so potentially requires a large number of terms for accuracy.
By optimizing this epoch, one can make a low order expansion as
accurate as the next higher order~\cite{Motohashi:2015hpa}.
This is especially advantageous in the EFT case where at each order there
are a multitude of slow roll parameters associated with the 5 fundamental
functions of time $H,b_s,c_s,b_t,c_t$.

\subsection{Optimized slow roll}%%%%%%%%%%%%%%%%%%%%%

In this section, we review the optimized slow-roll (OSR) approach developed systematically
by \cite{Motohashi:2015hpa} based on
earlier work in \cite{Stewart:2001cd}. 
If the temporal variations are sufficiently long, the  power spectrum  can
be approximated locally as a Taylor series around some fiducial $k$ which freezes out around some 
epoch $x_f$.
Given the integral formula  \eqref{intform}, we can relate this series to the Taylor
series of  the source function $G$ around 
$\ln x=\ln x_f$.  We can evaluate the integral formula \eqref{intform} term by term in the expansion to
obtain 
\begin{align} \label{lnD2} 
\ln \Delta^2 &\approx G(\ln x_f) + \sum_{p=1}^\infty q_p(\ln x_f) G^{(p)} (\ln x_f) , \notag\\
\f{d\ln \Delta^2}{d\ln k} &\approx - G'(\ln x_f) - \sum_{p=1}^\infty q_p(\ln x_f) G^{(p+1)} (\ln x_f), \notag\\
\alpha &\approx G''(\ln x_f) + \sum_{p=1}^\infty q_p(\ln x_f) G^{(p+2)} (\ln x_f) ,
\end{align} 
where we have used the fact that
\begin{equation}
\frac{d G^{(p)} (\ln x)}{d\ln k} = - G^{(p+1)}(\ln x).
\end{equation}
The coefficients $q_p(\ln x_f)$ are given by 
\begin{align} 
q_1(\ln x_f) &= \ln x_1 - \ln x_f ,\notag\\
\ln x_1 &\equiv \f{7}{3} - \ln 2 - \gamma_E ,
\end{align} 
and 
\begin{align} 
q_p(\ln x_f) &= \sum^p_{n=0} \f{c_{p-n}}{n!} q_1^n(\ln x_f) ,\notag\\
c_p &= \f{1}{p!} \lim_{z\to 0} \f{d^p}{dz^p} \kk{ e^{-z \mk{\f{7}{3}-\gamma_E} } \cos \mk{ \f{\pi z}{2} } \f{3\Gamma(2+z)}{(1-z)(3-z)}  } .
\end{align} 
Here, $\gamma_E$ is the Euler-Mascheroni constant.  
Specifically, $c_0=1, c_1=0, c_2=\f{4-3\pi}{72}, c_3=\f{55}{81}-\f{\zeta(3)}{3},\cdots$.
Note that the coefficients $q_p(\ln x_f)$ are the same for scalar and tensor perturbations, and do not depend on inflationary model, while they do depend on the choice of the evaluation epoch $x_f$.
For simplicity, we refer to the first terms of the right hand sides of \eqref{lnD2} as the leading order terms. 
We follow the usual conventions in defining the scalar and tensor
tilts as
\begin{align} 
n_s-1 & \equiv \frac{d\ln \Delta_\zeta^2}{d\ln k}, \notag\\
n_t & \equiv \frac{d\ln \Delta_{\gamma}^2}{d\ln k}.
\end{align}

For observationally viable models with $\Delta N>1$, the scalar tilt  $n_s-1= {\cal O}(1/N) \sim $ few percent. 
On the other hand, for the running of the tilt to be observable in the near future $\alpha_s = {\cal O}(1-n_s)$
and so these models violate the usual assumption that $\Delta N\sim N$.   
We therefore continue to assume $G' = {\cal O}(1/N)$ but take  $G^{(p+1)}/G^{(p)} \sim {\cal O}(1/\Delta N)$ 
where we allow $\Delta N \le N$.  In other words we assume that the function $G$ is composed of
features of width $\Delta N$ on top of a much larger smooth component that is responsible for
driving the remaining  $N\sim 60$ efolds of inflation.

For moderate widths, the above expansions will rapidly converge.  Indeed, since
\be \lim_{p\to\infty}\f{q_p}{q_{p-1}}=-\f{1}{2}, \ee
the convergence criterion is given by
\be \lim_{p\to\infty}\left| \f{G^{(p+1)}}{G^{(p)}} \right| < 2 . \ee
For $\Delta N < 1/2$, one needs to evaluate GSR integral formula \eqref{intform} on a case by case basis~\cite{Miranda:2015cea}.

Provided $\Delta N \gtrsim 1$, we can  truncate the series at some finite order to obtain approximate results. 
The leading-order approximation of the standard slow-roll approach corresponds to 
evaluating the expansion~\eqref{lnD2} at the sound horizon exit, i.e.\ $\ln x_f=0$, 
and truncating it at the leading order:
\begin{align} \label{ssr0} 
\ln\Delta^2 &\approx G(0) ,  \notag\\
\f{d \ln\Delta^2}{d \ln k} &\approx - G'(0) , \notag\\  
\alpha &\approx G''(0) .   
\end{align}  
Since the next-leading-order $p=1$ term has 
the coefficient $q_1(0)=1.06$ for $\ln x_f=0$, the correction for the leading-order slow roll approximation~\eqref{ssr0}  is suppressed by $1.06/\Delta N$ compared to the leading-order contribution.
For $\Delta N\sim N\sim 60$, the correction is sufficiently suppressed and hence the leading order approximation works well.  
However, if $\Delta N \sim$ a few, the correction is not highly suppressed.

To improve the truncation for moderately varying $G$, we can optimize the evaluation epoch $x_f$~\cite{Motohashi:2015hpa}. 
For the leading-order OSR approximation, we choose the evaluation epoch as $\ln x_f=\ln x_1$, which is a solution of $q_1(\ln x_1)=0$, so that the next-leading-order $p=1$ correction identically vanishes.
This yields 
\begin{align} \label{osr0}
\ln \Delta^2 &\approx G(\ln x_1) , \notag\\
\frac{d \ln \Delta^2}{d\ln k} &\approx -G'(\ln x_1) , \notag\\
\alpha  &\approx G''(\ln x_1) ,
\end{align}
While these expressions 
are as simple as the leading-order slow-roll approximation~\eqref{ssr0},  the change in the evaluation epoch $\ln x_f=\ln x_1$ provides
a large improvement in accuracy when $\Delta N \ll N$.
Since $\ln x_1 \approx 1.06$, this corresponds to evaluating the sources approximately 
$\sim 1$ efold before the sound horizon exit.
The correction to the truncation comes from the next-to-next-leading-order $p=2$, for which the coefficient is given by $q_2(\ln x_1) = c_2 \approx -0.36$.
Hence, compared to the leading-order term, the correction is suppressed by $0.36/\Delta N^2$.
For instance, for $\Delta N\sim 3$, the correction for the standard slow-roll~\eqref{ssr0} is given by $1.06/\Delta N \sim 0.35$ whereas for OSR~\eqref{osr0}, it is $0.36/\Delta N^2\sim 0.04$.

The same logic applies to a general $p$-th order OSR truncation~\cite{Motohashi:2015hpa}.
In this case we choose the evaluation epoch as $\ln x_f=\ln x_{p+1}$, which is a solution of $q_{p+1}(\ln x_{p+1})=0$, so that the next-order $p+1$ correction identically vanishes.  
The optimized evaluation then allows us to use the same expression of the formula as the $p$-th order truncation of the standard slow-roll, but with the accuracy of a $(p+1)$th order truncation.  We focus on the
leading-order OSR expansion~\eqref{osr0} in the following.

\subsection{EFT slow roll parameters}%%%%%%%%%%%%%%%%%%%%%

Now let us relate the Taylor expansions of the $G$ source functions for the scalars and tensors
 to those of the underlying
EFT functions $H,b_s,c_s,b_t,c_t$ all considered as functions of efolds $N$.   
We follow the Hubble slow roll parameter convention in the literature and define $\e_H = -\f{d\ln H}{dN}$ 
with the higher order derivatives given by the hierarchy 
\be
\d_1 \equiv \f{1}{2}\f{d\ln \e_H}{dN} - \e_H ,\quad 
\d_{p+1} \equiv \f{d\d_p}{dN} + \d_p(\d_1 - p\e_H).
\ee
For the scalar and tensor sound speeds, we define
\be
\sigma_{i,1} \equiv \f{d\ln c_i}{dN} , \quad
\sigma_{i,p+1} \equiv \f{d \sigma_{i,p}}{dN}  ,
\ee
and likewise for the normalization factor $b_i$
\be
\xi_{i,1} \equiv \f{d\ln b_i}{dN} , \quad
\xi_{i,p+1} \equiv \f{d \xi_{i,p}}{dN} ,
\ee
where $i=s,t$ and $p \geq 1$.

For each function there is a hierarchy of derivative parameters that match the $G^{(p)}$ expansion.
As discussed in the previous section, we assume $G' = {\cal O}(1/N)$ and $G^{(p+1)}/G^{(p)} \sim {\cal O}(1/\Delta N)$ which then sets the expectations
for the EFT slow roll parameters.    Hence we assume
\begin{align}
\label{counting}
\{G', \e_H,\delta_1,\sigma_{i,1},\xi_{i,1} \} &= \O \mk{\f{1}{N}} ,  \notag\\
\{ G^{(p+1)}, \d_{p+1}, \sigma_{i,p+1}, \xi_{i,p+1} \} &=  \O \mk{\f{1}{N \Delta N^{p}}}.
\end{align}
Note that $H$ is
special in that both it and its  derivative $\epsilon_H$ appear in the leading order scalar power spectrum
$\Delta_\zeta^2$; both $\e_H$ and $\delta_1$ appear in its derivative and so are ${\cal O}(1/N)$.

We can now establish the direct relationship between $G^{(p)}$ and the EFT slow roll parameters.  
We of course always keep the leading order expressions assuming \eqref{counting}.
For generality and to be able to also describe  $\alpha$ to leading order in the normal case
where $\Delta N\sim N$ we first expand expressions up to $\O(1/N^2)$, i.e.\ we keep $\O(1/N \Delta N^{p})$ terms but still drop $\O(1/N^2 \Delta N^{p})$ terms.  This also implies that the first order iteration in the
GSR approximation of \eqref{intform} suffices for $\O(1/N^2)$ expressions in $n$ and $\alpha$ but
not $\Delta^2$.

Since $G$ is taken to be a function of efolds of the sound horizon rather than the scale factor, we
also expand the conversion
\be \f{dN}{d\ln s_i} = -\f{aHs_i}{c_i} ,  \ee
around $N$ as
\be \f{aHs_i}{c_i} \approx 1+\e_H + \sigma_{i1} +  \sigma_{i2} + 2\e_H \sigma_{i1} + \e_H (3\e_H+2\d_1)  .  \ee

Therefore, using \eqref{fs} for the scalars,
\begin{align}
G &\approx \ln \mk{\f{H^2}{8\pi^2 b_s \e_H c_s}} 
- \f{10}{3}\e_H - \f{2}{3}\d_1 - \f{7}{3}\sigma_{s1} - \f{1}{3}\xi_{s1} 
- \f{8}{3} \sigma_{s2} \notag\\
&~~~ - \f{23}{3}\e_H^2 - \f{18}{3}\d_1\e_H - \f{11}{3}\e_H\sigma_{s1} - \f{1}{3}\e_H\xi_{s1} - \f{2}{3}\d_1\sigma_{s1} + \f{2}{3}\sigma_{s1}^2 - \f{1}{3}\sigma_{s1}\xi_{s1} , \notag\\
G' &\approx 4\e_H + 2\d_1 + \sigma_{s1} + \xi_{s1} 
+ \f{2}{3} \d_2 + \f{7}{3}\sigma_{s2} + \f{1}{3}\xi_{s2} \notag\\
&~~~ + \f{32}{3}\e_H^2 + \f{28}{3}\d_1\e_H - \f{2}{3} \d_1^2 + 5\e_H\sigma_{s1} + 2\d_1\sigma_{s1} + \sigma_{s1}^2 + \e_H\xi_{s1} + \sigma_{s1}\xi_{s1}  ,\notag\\ 
G'' &\approx - 2\d_2 - \sigma_{s2} - \xi_{s2}
- \f{2}{3}\d_3 - \f{7}{3}\sigma_{s3} - \f{1}{3}\xi_{s3} \notag\\
&~~~ - 8\e_H^2 - 10\e_H \d_1 + 2\d_1^2  , \notag\\ 
G^{(p)}&\approx (-1)^{p+1} \mk{ 2 \d_p + \sigma_{s,p} + \xi_{s,p} + \f{2}{3} \d_{p+1} + \f{7}{3} \sigma_{s,p+1} + \f{1}{3}\xi_{s,p+1} } , \quad (p\geq 3)  ,
\end{align}
and \eqref{ft} for the tensors,
\begin{align}
G &\approx \ln\mk{\f{H^2}{2 \pi^2 b_tc_t}} 
- \f{8}{3} \e_H - \f{7}{3} \sigma_{t1} - \f{1}{3}\xi_{t1} - \f{8}{3} \sigma_{t2} \notag\\
&~~~ - 7\e_H^2 - \f{16}{3} \d_{t1} \e_H - 3 \e_H \sigma_{t1} - \f{1}{3}\e_H\xi_{t1} + \f{2}{3} \sigma_{t1}^2 - \f{1}{3}\xi_{t1}\sigma_{t1} , \notag\\
G' &\approx 2\e_H  + \sigma_{t1} + \xi_{t1} + \f{7}{3}\sigma_{t2} + \f{1}{3}\xi_{t2}  \notag\\
&~~~ + \f{22}{3}\e_H^2 + \f{16}{3}\d_1\e_H + 3\e_H\sigma_{t1} + \sigma_{t1}^2  + \e_H\xi_{t1} + \sigma_{t1}\xi_{t1}  ,\notag\\ 
G'' &\approx - \sigma_{t2} - \xi_{t2} - \f{7}{3}\sigma_{t3} - \f{1}{3}\xi_{t3} \notag\\
&~~~ - 4\e_H^2 - 4\e_H \d_1 , \notag\\ 
G^{(p)}&\approx (-1)^{p+1} \mk{ \sigma_{t,p} + \xi_{t,p} + \f{7}{3} \sigma_{t,p+1} + \f{1}{3}\xi_{t,p+1} } , \quad (p\geq 3)  ,
\end{align}
which recovers the result in \cite{Motohashi:2015hpa} for $c_i=b_i=1$ since $\sigma_{i,p}=\xi_{i,p}= 0$.

With these expressions we can explicitly give the parameters of the power spectrum to leading order
in the optimized slow roll approximation as
\begin{align}
\ln \Delta_\zeta^2 
&\approx \ln\mk{\f{H^2}{8 \pi^2 b_s c_s \e_H}} - \f{10}{3}\e_H - \f{2}{3}\d_1 - \f{7}{3}\sigma_{s1} - \f{1}{3}\xi_{s1} \Big|_{x=x_1}, \notag\\
n_s-1  
&\approx - 4\e_H - 2\d_1 - \sigma_{s1} - \xi_{s1} 
- \f{2}{3} \d_2 - \f{7}{3}\sigma_{s2} - \f{1}{3}\xi_{s2} \Big|_{x=x_1} , \notag\\
\alpha_s  
&\approx - 2\d_2 - \sigma_{s2} - \xi_{s2} - \f{2}{3}\d_3 - \f{7}{3}\sigma_{s3} - \f{1}{3}\xi_{s3} 
- 8\e_H^2 - 10\e_H \d_1 + 2\d_1^2 \Big|_{x=x_1} ,
\label{pss}
\end{align}
for scalars, and
\begin{align}
\ln \Delta_{\gamma}^2 
&\approx \ln\mk{\f{H^2}{2 \pi^2 b_tc_t}} 
- \f{8}{3} \e_H - \f{7}{3} \sigma_{t1} - \f{1}{3}\xi_{t1}  \Big|_{x=x_1}, \notag\\
n_t  
&\approx - 2\e_H - \sigma_{t1} - \xi_{t1} - \f{7}{3}\sigma_{t2} - \f{1}{3}\xi_{t2}  \Big|_{x=x_1} , \notag\\
\alpha_t 
&\approx - \sigma_{t2} - \xi_{t2} - \f{7}{3}\sigma_{t3} - \f{1}{3}\xi_{t3} 
- 4\e_H^2 - 4\e_H \d_1 \Big|_{x=x_1},
\label{pst}
\end{align}
for tensors. Here,
the right hand sides are evaluated at the optimized point $\ln x=\ln x_1\approx 1.06$ and
we have kept ${\cal O}(1/N^2)$ terms only for the running of the tilt parameters
since they are leading order if $\Delta N \approx N$.
Unlike the $P(X,\phi)$ inflation case, it is possible to have $n_T>0$ without having
$\epsilon_H < 0$ or growing $H$.  This would require  negative contributions from $\sigma_{t1},\xi_{t1},\sigma_{t2},\xi_{t2}$ that compensate $- 2\e_H$.

Finally note that there is a subtlety that must be kept in mind when comparing the
scalar and tensor spectra.  
Although both the scalar and the tensor parameters are evaluated at $x=x_1$, 
they represent different epochs during inflation, 
$s_s(N_{s}) = x_1/k$ and $s_t(N_{t})= x_1/k$ where $N_{s}\ne N_{t}$ if the sound speeds differ.   
Thus when combining these relations to form the tensor-to-scalar ratio at a fixed $k$, 
we must evaluate the common slow roll parameters at different epochs.  
Likewise the consistency relation
\begin{align}
r\equiv \f{4 \Delta_{\gamma}^2}{\Delta_\zeta^2}
&\approx 16\e_H \f{b_sc_s}{b_tc_t}\approx  - \f{8b_sc_s}{b_tc_t} n_t ,
\label{tsr}
\end{align}
only applies when $b_s,c_s,b_t,c_t$ are exactly constant even at leading order 
(see Eq.~(4.43) in \cite{Kobayashi:2011nu} for Horndeski theory).  
More generally, one would use \eqref{pss} evaluated at $k s_s=x_1$ and
\eqref{pst} evaluated at $k s_t=x_1$ which does not provide a strict consistency
relationship between the $r$ and $n_T$ observables.

To summarize, the expressions \eqref{pss} and \eqref{pst} apply to any inflationary model that has the quadratic actions \eqref{S2s} and \eqref{S2t} with the standard dispersion relation, so long as 
the scalar and tensor perturbations freezeout after crossing their respective sound horizons and
the sources $G_\zeta$ and $G_\gamma$ are moderately slowly varying with $\Delta N > 1$.
Given a specific Lagrangian, one could check the above conditions, and then calculate $H,b_s,c_s,b_t,c_t$ and their slow-roll parameters to obtain power spectra.  The correction to the truncation is suppressed by $0.36/\Delta N^2$ in contrast with $1/\Delta N$ suppression for the standard slow-roll leading-order approximation.

\section{Conclusion}%%%%%%%%%%%%%%%%%%%%%%%%%%%%%%%%%%%%%%%%%
\label{sec:con}

We have unified and streamlined the calculation of scalar and tensor power spectra observables
in the EFT of inflation using its ADM form.  
The subset 
that describes theories that have only second-order spatial derivatives in the EOMs for their
perturbations leads to a quadratic action for scalar perturbation~\eqref{S2s} and tensor perturbation~\eqref{S2t}
with normal dispersion relations.  This class includes Horndeski and GLPV theories as well as their
canonical and $P(X,\phi)$ subsets.
The evolution of the scalar and tensor perturbations is characterized by 4 free functions of time $b_s,c_s,b_t,c_t$ in addition to the usual background expansion rate $H$.  
The information in these functions can be further condensed into 2 sources for the scalar
and tensor power spectra $G_\zeta$ and  $G_\gamma$ that are functions of the 2 respective sound horizons.

We give the criteria under which scalar and tensor perturbations freeze out after crossing their respective
sound horizon and under which the unitary and comoving gauge coincide in the scalar curvature in Appendix \ref{sec:gau}.  
In this case, we utilize  the generalized slow roll approach to obtain an integral expression for their power spectra~\eqref{intform},
assuming small, but not necessarily slowly varying deviations from scale invariance in the 2 source functions.
For cases when variations occur on the efold time scale or slower, we provide explicit expressions in terms
of  5 slow-roll hierarchies of parameters for $b_s,c_s,b_t,c_t$ and $H$.   By optimizing the evaluation of these slow 
roll parameters, we greatly improve the accuracy of the truncated hierarchies leading to simple but accurate expressions in
terms of leading order parameters.

\acknowledgments%%%%%%%%%%%%%%%%%%%%%%%%%%%%%%%%%%%%%%%%%
This work was supported by the Kavli Institute for Cosmological Physics at the University of Chicago through grants NSF PHY-0114422 and NSF PHY-0551142 and an endowment from the Kavli Foundation and its founder Fred Kavli.  
H.M.\ was supported in part by MINECO Grant SEV-2014-0398,
PROMETEO II/2014/050,
Spanish Grant FPA2014-57816-P of the MINECO, and
European Union’s Horizon 2020 research and innovation programme under the Marie Sk\l{}odowska-Curie grant agreements No.~690575 and 674896.
H.M.\ thanks the Research Center for the Early Universe, where part of this work was completed.  
W.H.\  was additionally supported by  U.S.~Dept.\ of Energy
contract DE-FG02-13ER41958 and NASA ATP NNX15AK22G
and
thanks the Aspen Center for Physics, which is supported by National Science Foundation grant PHY-1066293, 
where part of this work was completed.

%%%%%%%%%%%%%%%%%%%
\appendix
%%%%%%%%%%%%%%%%%%%

\section{Relationship to Literature}%%%%%%%%%%%%%%%%%%%%%%%%%%%%%%%%%%%%%%%%%
\label{sec:dic}

In this section, we present correspondence between our notation and that in the literature.
We also highlight the advantages of our analysis for the EFT action~\eqref{LagEFT}, notation of~\eqref{cxynot}, and
simplicity of the quadratic Lagrangian \eqref{LagEFT4} in comparison.

\subsection{Gleyzes, Langlois, Piazza, \& Vernizzi (2013)}%%%%%%%%%%%%%%%%%%%%%

Gleyzes, Langlois, Piazza, \& Vernizzi~\cite{Gleyzes:2013ooa} study the Lagrangian 
\be \label{LagpastEFT} S = \int d^4x N\sqrt{h} L(N,K,R,\mS,\Z,\Y;t),  \ee 
where 
\be
\mS \equiv K^i_{\hphantom{i}j} K^j_{\hphantom{i}i} , \quad 
\Z \equiv R^i_{\hphantom{i}j} R^j_{\hphantom{i}i} , \quad
\Y \equiv R^i_{\hphantom{i}j} K^j_{\hphantom{i}i} ,
\ee
which is a subset of \eqref{LagEFT} that is equivalent at the level of the quadratic action.
The perturbations of these combinations around the flat FLRW metric are given by 
\be \label{dSdZ}
\delta \mS = 2H \delta K + \delta K^i_{\hphantom{i}j} \delta K^j_{\hphantom{i}i}  ,\quad 
\delta \Z = \delta R^i_{\hphantom{i}j} \delta R^j_{\hphantom{i}i}  , \quad
\delta \Y = H\delta R + \delta K^i_{\hphantom{i}j} \delta R^j_{\hphantom{i}i} ,
\ee
which mixes the structure of the quadratic Lagrangian. 
For example, the linear $\delta R$ term is given by 
\be L \supset L_{,R} \delta R + L_{,\U} \delta \U = (L_{,R} + H L_{,\U}) \delta R, \ee
where $L_{,Y} \equiv \f{\pa L }{\pa Y} |_{\rm b}$ is evaluated at the background in their notation.
In our notation, the $\delta R$ term is simply given by $\C_R\delta R$ in \eqref{LagEFT2}.
As an example at quadratic order, the $\delta K^2$ term is given by 
\begin{align} 
L &\supset \f{1}{2} L_{,KK} \delta K^2 + L_{,\mS K} \delta \mS \delta K + \f{1}{2} L_{,\mS \mS} \delta \mS^2 \notag\\
&= \f{1}{2} ( L_{,KK} + 4 H L_{,\mS K} + 4 H^2 L_{,\mS \mS} )  \delta K^2 ,
\end{align} 
whereas in our notation the entire term is represented by $\f{1}{2} \C_{KK} \delta K^2$ in \eqref{LagEFT2}.
Our notation makes the correspondence between the EFT coefficients of the quadratic Lagrangian
and the EFT Lagrangian transparent.

Beyond the above notational difference, they performed an additional integration by parts of the $\delta K^i_{\hphantom{i}j} \delta R^j_{\hphantom{i}i}$ term 
so as to rewrite the $\Y$ dependence in terms of the other 
existing $N,R,K,\mS$ terms and reduce the total number of EFT coefficients.
Specifically they exploit
\be \label{dKdRint} N\sqrt{h} \tilde\C_{KR} \delta K^i_j \delta R^j_i \sim \f{a^3}{2} \kk{ (\dot{\tilde\C}_{KR} + H\tilde\C_{KR})\mk{\f{\delta \sqrt{h}}{a^3} \delta R + \delta_2 R } + \tilde\C_{KR} \delta R\delta K + H\tilde\C_{KR} \delta N \delta R } . \ee 
Again one can obtain the more transparent form~\eqref{LagEFT2} by omitting the process.

To fully translate from the notation of \cite{Gleyzes:2013ooa}, we have
\begin{align}
\bar L &= \C ,\notag\\
L_{,N} &= \C_N, \notag\\
L_{,NN} &= \C_{NN} ,\notag\\
L_{,\mS} &= \f{1}{2} \tilde\C_{KK} , \notag\\
L_{,\Z} &= \f{1}{2} \tilde \C_{RR} , \notag\\
{\cal A} &= \C_{KK} , \nonumber\\
{\cal B} &= \C_{NK} , \nonumber\\
{\cal F} &= \C_{K} ,\notag\\
L_{,R} + \f{1}{2} \dot L_{,\Y} + \f{3}{2} H L_\Y &= \C_{R} + \frac{1}{2} \dot{\tilde \C}_{KR} + \frac{1}{2} H\tilde \C_{KR} ,\notag\\
L_{,RR} + H^2L_{,\Y\Y} + 2HL_{,\Y R} &= \C_{RR} ,\notag\\ 
L_{,NR} + H L_{,N\Y} - \f{1}{2} \dot L_{,\Y} &= \C_{NR} - \frac{1}{2} \dot{\tilde \C}_{KR} ,\notag\\
\C + HL_{K\Y} + 2H^2L_{\mS\Y} + \f{1}{2}L_{\Y} &= \C_{KR} + \frac{1}{2} \tilde \C_{KR} .
\end{align}
The notation~\eqref{cxynot} simplifies the coefficients as can be seen in the right hand sides.  
Note that the right hand sides have additional $\dot{\tilde \C}_{KR}$ and $\tilde \C_{KR}$ terms, which come from additional integration by parts of the $\delta K^i_{\hphantom{i}j} \delta R^j_{\hphantom{i}i}$ term.
One can confirm that their quadratic Lagrangian in Eq.~(21) with the definition Eqs.~(13) and (127) of \cite{Gleyzes:2013ooa} and our \eqref{LagEFT4} are equivalent up to total derivative after using these correspondences and the identity~\eqref{dKdRint}.
We thus obtain the same scalar, vector, and tensor equations of motion.\footnote{In their intermediate equations
Eqs.~(12)-(21) there are additional terms 
$\dot \F+L_{,N} = \dot \C_K + \C_N$, which vanish by virtue of background equations~\eqref{eqn:background}.
}

\subsection{Kase \& Tsujikawa (2015)}%%%%%%%%%%%%%%%%%%%%%

Kase \& Tsujikawa~\cite{Kase:2014cwa} extended the above approach by adding additional dependencies to the Lagrangian \eqref{LagpastEFT} on new types of combinations, which include spatial covariant derivatives such as $\nabla_i R \nabla^i R$, and/or the acceleration $a_\mu\equiv n^\nu n_{\nu;\mu}$ such as $a_i a^i$.
While the Lagrangian~\eqref{LagEFT} does not include these types of combinations it is contained as a subset within which we can establish the correspondences.

To translate from the notation of \cite{Kase:2014cwa}, we have
\begin{align}
\bar L &= \C ,\notag\\
L_{,N} &= \C_N, \notag\\
L_{,NN} &= \C_{NN} ,\notag\\
L_{,\mS} &= \f{1}{2} \tilde\C_{KK} , \notag\\
L_{,\Z} &= \f{1}{2} \tilde \C_{RR} , \notag\\
{\cal A} &= \C_{KK} , \nonumber\\
{\cal B} &= \C_{NK} , \nonumber\\
{\cal C} &= \C_{KR} +\frac{1}{2} \tilde \C_{KR} , \nonumber\\
{\cal D} &= \C_{NR} - \frac{1}{2} \dot{\tilde \C}_{KR} , \nonumber\\
{\cal E} &= \C_{R} + \frac{1}{2} \dot{\tilde \C}_{KR} + \frac{1}{2} H\tilde \C_{KR} , \nonumber\\
{\cal F} &= \C_{K} ,\notag\\
{\cal G} &= \C_{RR} .
\end{align}
Again, with these correspondences and the identity~\eqref{dKdRint}, their quadratic Lagrangian in Eq.~(4.29) matches our \eqref{LagEFT4}.  
We thus obtain the same scalar, vector, and tensor equations of motion.

For the restriction to theories having up to second order spatial derivatives in EOMs, the scalar quadratic Lagrangian is given as their Eq.~(4.59), which matches our \eqref{S2s} with
\begin{eqnarray}
{\cal W} &=& \C_{NK} -2 H \C_{KK} , \nonumber\\
{\cal M} &=& \f{1}{2} {\cal A}_{\dot\zeta \zeta} - \C_{KR} , \nonumber\\
Q_s &=& {\cal A}_{\dot\zeta\dot\zeta} .
\end{eqnarray}

\subsection{Gleyzes, Langlois, \& Vernizzi (2015)}%%%%%%%%%%%%%%%%%%%%%

Gleyzes, Langlois, \& Vernizzi~\cite{Gleyzes:2014rba} also extend 
our Lagrangian \eqref{LagEFT} to cases where there are extra spatially covariant derivatives
similar to \cite{Kase:2014cwa} and introduced the tensor derivative structures for
$K^i_{\hphantom{i}j}$ and $R^i_{\hphantom{i}j}$ that we generalize in
\eqref{cxynot}. Again we can compare results for the subset that omits these additions.

Aside from compactness of notation, the conceptual difference with our treatment is again  that they performed additional integration by parts of $\delta K^i_{\hphantom{i}j} \delta R^j_{\hphantom{i}i}$, which we discuss in \eqref{dKdRint} above.   
However in this case we could not establish agreement in the final quadratic Lagrangian due to what are apparently typos in the current
arXiv:1411.3712v2~\cite{Gleyzes:2014rba}
\begin{align} \label{crn}
{\rm Eq.~(55):}~~~ & \C \delta K^i_j \delta R^j_i \to 2\C \delta K^i_j \delta R^j_i ,\notag\\
{\rm Eq.~(60):}~~~ & \G^* = \G + \dot\C + H \C ,\notag\\ 
 & \C^* = \hat \C + \C ,\notag\\  
 & \B_R^* = \B_R - \dot \C ,
\end{align}
where we used $\bar N=1$ in comparison with their original expressions.  
Their subsequent equations and quantities based on the above variables such as $c_T,\alpha_T$ in \cite{Gleyzes:2014rba} 
should be corrected with this relation~\eqref{crn}.  
Some but not all of these typos are addressed in 
\cite{Gleyzes:2015pma} and consequently those that require correction in  \cite{Gleyzes:2015pma}  include
\be {\rm Eq.~(2.25):}~~~ \alpha_T \equiv \f{\G + \dot\C + H \C}{\A_K} - 1 . \ee

To translate the notation of \cite{Gleyzes:2014rba}, including the corrections of \eqref{crn},
we have
\begin{align} \label{dic3}
\bar L &= \C ,\notag\\
L_{,N} &= \C_N, \notag\\
L_{,NN} &= \C_{NN} ,\notag\\
\hat {\cal A}_K &= \C_{KK} , \nonumber\\
{\cal A}_K &= \frac{1}{2} \tilde \C_{KK} , \nonumber\\
\hat {\cal A}_R &= \C_{RR} , \nonumber\\
{\cal A}_R &= \f{1}{2}\tilde \C_{RR} , \nonumber\\
{\cal B} &= \C_{NK} , \nonumber\\
{\cal B}_{R}^* &= \C_{NR} - \frac{1}{2} \dot {  \tilde \C} _{KR} , \nonumber\\
\C &= \f{1}{2}\tilde\C_{KR} , \notag\\
{\cal G}^* &= \C_R +\frac{1}{2} \dot{\tilde \C} _{KR} +\frac{1}{2} H  \tilde \C_{KR} , \nonumber \\
\hat {\cal C} &= \C_{KR} , \nonumber\\
\hat {\cal C}^*  &= \C_{KR}+ \frac{1}{2}\tilde \C_{KR} .
\end{align}
With the corrections~\eqref{crn}, the above correspondences~\eqref{dic3}, and the identity~\eqref{dKdRint}, their quadratic Lagrangian in Eq.~(59) matches our \eqref{LagEFT4}.

From \eqref{dic3} we also have\footnote{$\alpha_M,\alpha_K,\alpha_B,\alpha_T$ were introduced in \cite{Bellini:2014fua} but with a different normalization for $\alpha_B=-\C_{NK}/H \tilde \C_{KK}$.}
\begin{align} \label{dic4}
M^2 &= \tilde \C_{KK} ,\notag\\
\alpha_M &= \f{1}{H} \f{d}{dt} \ln \tilde \C_{KK} ,\notag\\
\alpha_K &= \f{2\C_N+\C_{NN}}{H^2\tilde\C_{KK}} ,\notag\\
\alpha_B &= \f{\C_{NK}}{2H\tilde\C_{KK}} ,\notag\\
\alpha_T &= \f{2 \C_R + \dot {\tilde \C} _{KR} + H  \tilde \C_{KR}}{\tilde \C_{KK}} - 1 ,\notag\\
\alpha_H &= \f{2 \C_R + 2\C_{NR} + H  \tilde \C_{KR}}{ \tilde \C_{KK}} - 1  .
\end{align}
With these correspondences, the tensor quadratic Lagrangian in Eq.~(66) of \cite{Gleyzes:2014rba} 
matches our \eqref{S2t-1} up to total derivative.
For the scalar quadratic Lagrangian, with the correspondences~\eqref{dic4}, the assumption~\eqref{eqn:conditions}, and noting that 
\be \f{1+\alpha_H}{1+\alpha_B} = \f{H}{\C_{KK}}( 2\C_{KR} - \A_{\dot\zeta\zeta} ) , \ee
we have
\begin{align}
\L_{\dot\zeta\dot\zeta} &= 2\A_{\dot\zeta\dot\zeta},\notag\\
\L_{\pa\zeta\pa\zeta} &= -2\e_H b_s,
\end{align}
and thus the scalar sector of Eq.~(79) of \cite{Gleyzes:2014rba} matches our \eqref{S2s}.

\section{Unitary vs Comoving Gauge}%%%%%%%%%%%%%%%%%%%%%%%%%%%%%%%%%%%%%%%%%
\label{sec:gau}

While tensor perturbations are gauge invariant, the scalar curvature perturbations are not.  Hence the question of which curvature spectrum controls observable quantities arises.
In this Appendix, we clarify the difference between curvature perturbations in the unitary gauge, 
used in the main text, and the comoving gauge used in initial conditions for evolving the observables after inflation.

Let us first consider the most general description of scalar perturbations in a mode with wavenumber
$k$ around the flat FLRW metric
\be \label{fullds2} ds^2 =  - (1+2AQ) dt^2 + 2 a B Q_i dt dx^i + a^2(\d_{ij} + 2H_L Q\d_{ij} + 2H_T Q_{ij}) dx^idx^j , \ee
where $Q$ is an eigenfunction of the Laplace operator $\delta^{ij} \partial_i\partial_j Q = -k^2Q$ and 
\begin{align}
Q_i &= -k^{-1} \pa_i Q, \notag\\
Q_{ij} &= \mk{k^{-2} \pa_i\pa_j + \f{1}{3} \d_{ij}} Q .
\end{align}
In the spatially flat background assumed here $Q$ are simply plane waves.
The metric fluctuations transform under a diffeomorphism or gauge transformation
 $x^\mu \to x^\mu + \epsilon^\mu$ with $\epsilon^0= T Q$ and $\epsilon^i= L \delta^{ij} Q_j$
 as
 \begin{align} 
\delta_\e\mk{  H_L + \f{ H_T}{3} } &= - HT ,  \notag\\
\delta_\e ( A ) &= - \dot T ,\notag\\
\delta_\e (B )&=  a \dot L + \frac{k}{a} T, \notag\\
\delta_\e (H_T) &= k L,
\end{align}
where $H_L+H_T/3$ is the curvature perturbation.
Unitary and comoving gauges correspond to placing conditions on the metric fluctuations $A,B,H_L,H_T$ that
fix this gauge freedom.
For the comoving gauge condition, it is useful to note that the $0i$ perturbation to the
Einstein tensor
\be \delta  G^0_{\hphantom{0}i}= \G_v Q_i \ee
is given by 
\be \label{Gvgen}
\G_v = \frac{2k}{a} \left[ HA - \mk{ \dot H_L + \f{1}{3}\dot H_T } \right] .  \ee
This combination transforms under a gauge transformation as
\be \delta_\e \G_v = \frac{2k}{a} \dot HT . \ee

\subsection{Comoving gauge}%%%%%%%%%%%%%%%%%%%%%

The curvature perturbation in comoving gauge is usually taken as the initial conditions  from inflation for structure
formation.  Comoving gauge is so named because 
for  canonical inflation, the perturbed Einstein equation is given by $\delta G_{\mu\nu} = \delta T^{\phi}{}_{\mu\nu}$.   Comoving time slicing is defined by the vanishing of the the momentum 
density associated with field perturbations $\delta T^{\phi}{}^0_{\hphantom{0}i}=0$.  
We can generalize this treatment to cases where the Einstein equation does not hold by defining
comoving slicing such that
 $\delta G^0_{\hphantom{0}i}=0$ (see \cite{Hu:2016wfa}).    This condition sets $\G_v=0$ and
 completely specifies 
 the time slicing $T$ whereas setting $H_T=0$ completely fixes the spatial gauge freedom.
Provided that the scalar field decays into matter after inflation, the condition $\delta G^0_{\hphantom{0}i}=0$ will be smoothly connected to the usual comoving gauge condition $\delta T^{{\rm m}}{}^0_{\hphantom{0}i} =0$ used as the initial conditions for structure formation.

To avoid confusion, we denote the curvature perturbation and the lapse in the comoving gauge as
\be \curv = H_L+\f{H_T}{3}, \quad \xi = A . \ee  
It is shown in \cite{Hu:2016wfa} that since $\dot \curv = \f{\dot a}{a} \xi$, for any metric theory
\be \label{lapsecondition} |\xi| \ll |\curv| ~~\Longrightarrow~~ \left|\f{1}{\curv}\f{d\curv}{dN}\right| \ll 1 , \ee
if the background spatial curvature vanishes.  Hence when the lapse is much smaller than the curvature,
the curvature is approximately conserved on the efold time scale.
The comoving gauge lapse function is given by
\be \xi = -\f{\delta p}{\rho+p} + \f{2}{3}\frac{ p\pi}{\rho+p}  , \ee
where $\rho,p,\pi$ are the components of $G_{\mu\nu}$ that would
be associated with total energy density, pressure, and anisotropic stress given
the Einstein equations (see \cite{Hu:2016wfa} for details).   Note that in this definition $\rho+p = - dH^2/dN = 2 \epsilon_H H^2$.

Therefore, the condition \eqref{lapsecondition} can be violated even outside the horizon when $dH^2/dN \rightarrow 0$, which happens in ultra slow-roll inflation with a canonical kinetic term \cite{Kinney:2005vj,Namjoo:2012aa,Martin:2012pe,Motohashi:2014ppa,Motohashi:2017aob}, certain $P(\phi,X)$ models \cite{Chen:2013aj,Huang:2013lda,Chen:2013eea}, and certain Horndeski models through the $\Box \phi$ term \cite{Hirano:2016gmv}.   In these
cases superhorizon fluctuations cannot be absorbed into a separate universe construction
and hence violate non-Gaussianity consistency relations.

\subsection{Unitary gauge}%%%%%%%%%%%%%%%%%%%%%

While the EFT Lagrangian \eqref{LagEFT} does not depend on the scalar field $\phi$ and so 
the quadratic Lagrangian does not include $\d\phi$, we can regard it as the  quadratic action in the unitary gauge.  The gauge transformation on the scalar field acts as $\delta_e (\d \phi) = -\dot\phi T$ and
so the unitary gauge condition $\d\phi=0$ completely fixes the time slicing.  

Unitary gauge is employed in the EFT of inflation so as to
express the dynamical degrees of freedom through the metric alone.
In addition, it is often employed for analysis of scalar-tensor theories involving nontrivial derivative couplings, e.g.\ GLPV and Horndeski theories \cite{Kobayashi:2011nu}. 
This is because the unitary gauge fixing condition simplifies calculation by dropping derivative terms of $\delta\phi$ which are present in \eqref{GLPV}, yielding a scalar quadratic action for the metric
degrees of freedom in the standard form \eqref{S2s}. 
While unitary gauge coincides with  comoving gauge for the $P(\phi,X)$ model
\eqref{PXp}, they do not for general scalar-tensor theories.

In terms of the single $k$-mode representation of \eqref{fullds2}, unitary gauge sets
\begin{align}
\alpha = A,\quad
\beta = B, \quad
\zeta = H_L,\quad
0 = H_T .
\end{align}
These harmonic  amplitudes are related to the spatial metric fluctuations of \eqref{unimet}
by $\delta N \rightarrow \alpha Q$, $\partial_i \psi \rightarrow a  \beta Q_i$, $\zeta \rightarrow \zeta Q$.

In  unitary gauge,
$\G_v$ from \eqref{Gvgen} reduces to 
\be \G_v = \f{2k}{a} \Delta \ee
where
\be \label{defDelta} \Delta \equiv H\alpha - \dot \zeta. \ee
Using \eqref{Gvgen} we can define the time shift from unitary gauge to comoving gauge as
\begin{equation} \label{Tuc}
T = -\frac{ a }{2k\dot H} \G_v = -\frac{\Delta}{\dot H} = \frac{\Delta}{\epsilon_H H^2}
\end{equation}
and thus
\begin{align} \label{unicom}
\curv &= \zeta - \f{\Delta}{H\e_H} , \notag\\
\xi &= \alpha - \f{d}{dt} \mk{ \f{\Delta}{H^2\e_H}  } .
\end{align}
Therefore, the conservation of $\zeta$ is not strictly equivalent to $\curv$ and the two curvatures need not coincide.
We next consider the conditions under which the two do coincide.

\subsection{Curvature equivalence}%%%%%%%%%%%%%%%%%%%%%

If the contribution of $\Delta$ is negligible in \eqref{unicom}, the curvature perturbation in  unitary gauge coincides with that in  comoving gauge.
For the  theories considered here, i.e.~those possessing second-order EOMs for scalar perturbations, including Horndeski and GLPV theories, we can use the constraint equation \eqref{coneq-lapse} for the lapse to obtain 
\be \Delta = \Gamma \dot\zeta, \ee
where 
\be \label{Gam} \Gamma \equiv  {  \f{\C_{NK}}{2H\C_{KK}-\C_{NK}}  } . \ee
As expected, for the canonical case~\eqref{eqn:canonical}, $\Gamma = 0$. 
Note that $\Gamma$ diverges if the condition \eqref{dencond} is violated.     We can trace the origin of this divergence to an infinite time shift $T$ between unitary and comoving gauges.

From \eqref{unicom}, the comoving curvature coincides with the unitary curvature when
\begin{equation}
\left|  \frac{d\ln \zeta}{dN } \right|   \ll  \left| \frac{\epsilon_H}{\Gamma}\right| .
\end{equation}
There are two possible cases which satisfy this condition.
The first case is 
\be \label{cond1} \Gamma \approx 0, \ee
which means that the model of interest is very close to the canonical inflation. 
The second case is when the unitary gauge curvature is nearly constant
\begin{equation}
\frac{d\ln \zeta}{dN }\approx 0 .
\end{equation}
We have already seen in \eqref{shsol-s} that conservation of $\zeta$ above the sound horizon
requires that 
\be \label{cond2}
\f{c_s^2}{a^3b_s\e_H} \propto s_s^p. \ee
with $p(s_s)>0$.   Note that even if \eqref{cond1} is satisfied such that ${\cal R} \approx \zeta$,
\eqref{cond2} must also be satisfied in order to have ${\cal R} \approx $\,const. above the
sound horizon.

If \eqref{cond2} is satisfied then even if $\epsilon_H/\Gamma$ is finite, the two curvatures
will eventually coincide as $\eta \rightarrow 0$, but potentially not until well after sound horizon
crossing.   More concretely, if 
\be
\frac{d\ln \zeta}{dN } \approx (ks_s)^p,
\ee
the difference between the two gauges reads
\begin{align}
\curv &\approx \zeta \mk{1 - \f{\Gamma (ks_s)^p}{\e_H} } , \notag\\
\xi &\approx \alpha - \f{d}{dt} \mk{ \f{\Gamma \zeta (ks_s)^p}{H\e_H} } .
\end{align}
Therefore, sufficiently after sound horizon crossing
\be (ks_s)^p \ll \min\left(\left| \f{\e_H}{\Gamma}\right|,1\right), \ee
one can approximate $\curv \approx \zeta$ and $\xi \approx \alpha$.

In conclusion, if \eqref{cond2} is satisfied [regardless of whether \eqref{cond1} is satisfied],
one can calculate the power spectrum  $\Delta_\zeta^2$ as described in the main text.  Since
$\Delta_\zeta^2$ is constant outside the sound horizon its freezeout value is the same as
its value at $ks_s \rightarrow 0$.   Hence we can take $\Delta_{\cal R}^2=\Delta_\zeta^2$ so
long as \eqref{cond2} is satisfied between freezeout and the epoch at which we
evaluate $\Delta_{\cal R}^2$.

\bibliography{ref-HornInf}

\begin{thebibliography}{55}%
\makeatletter
\providecommand \@ifxundefined [1]{%
 \@ifx{#1\undefined}
}%
\providecommand \@ifnum [1]{%
 \ifnum #1\expandafter \@firstoftwo
 \else \expandafter \@secondoftwo
 \fi
}%
\providecommand \@ifx [1]{%
 \ifx #1\expandafter \@firstoftwo
 \else \expandafter \@secondoftwo
 \fi
}%
\providecommand \natexlab [1]{#1}%
\providecommand \enquote  [1]{``#1''}%
\providecommand \bibnamefont  [1]{#1}%
\providecommand \bibfnamefont [1]{#1}%
\providecommand \citenamefont [1]{#1}%
\providecommand \href@noop [0]{\@secondoftwo}%
\providecommand \href [0]{\begingroup \@sanitize@url \@href}%
\providecommand \@href[1]{\@@startlink{#1}\@@href}%
\providecommand \@@href[1]{\endgroup#1\@@endlink}%
\providecommand \@sanitize@url [0]{\catcode `\\12\catcode `\$12\catcode
  `\&12\catcode `\#12\catcode `\^12\catcode `\_12\catcode `\%12\relax}%
\providecommand \@@startlink[1]{}%
\providecommand \@@endlink[0]{}%
\providecommand \url  [0]{\begingroup\@sanitize@url \@url }%
\providecommand \@url [1]{\endgroup\@href {#1}{\urlprefix }}%
\providecommand \urlprefix  [0]{URL }%
\providecommand \Eprint [0]{\href }%
\providecommand \doibase [0]{http://dx.doi.org/}%
\providecommand \selectlanguage [0]{\@gobble}%
\providecommand \bibinfo  [0]{\@secondoftwo}%
\providecommand \bibfield  [0]{\@secondoftwo}%
\providecommand \translation [1]{[#1]}%
\providecommand \BibitemOpen [0]{}%
\providecommand \bibitemStop [0]{}%
\providecommand \bibitemNoStop [0]{.\EOS\space}%
\providecommand \EOS [0]{\spacefactor3000\relax}%
\providecommand \BibitemShut  [1]{\csname bibitem#1\endcsname}%
\let\auto@bib@innerbib\@empty
%</preamble>
\bibitem [{\citenamefont {Creminelli}\ \emph {et~al.}(2006)\citenamefont
  {Creminelli}, \citenamefont {Luty}, \citenamefont {Nicolis},\ and\
  \citenamefont {Senatore}}]{Creminelli:2006xe}%
  \BibitemOpen
  \bibfield  {author} {\bibinfo {author} {\bibfnamefont {P.}~\bibnamefont
  {Creminelli}}, \bibinfo {author} {\bibfnamefont {M.~A.}\ \bibnamefont
  {Luty}}, \bibinfo {author} {\bibfnamefont {A.}~\bibnamefont {Nicolis}}, \
  and\ \bibinfo {author} {\bibfnamefont {L.}~\bibnamefont {Senatore}},\ }\href
  {\doibase 10.1088/1126-6708/2006/12/080} {\bibfield  {journal} {\bibinfo
  {journal} {JHEP}\ }\textbf {\bibinfo {volume} {12}},\ \bibinfo {pages} {080}
  (\bibinfo {year} {2006})},\ \Eprint {http://arxiv.org/abs/hep-th/0606090}
  {arXiv:hep-th/0606090 [hep-th]} \BibitemShut {NoStop}%
%%CITATION = HEP-TH/0606090;%%
\bibitem [{\citenamefont {Cheung}\ \emph {et~al.}(2008)\citenamefont {Cheung},
  \citenamefont {Creminelli}, \citenamefont {Fitzpatrick}, \citenamefont
  {Kaplan},\ and\ \citenamefont {Senatore}}]{Cheung:2007st}%
  \BibitemOpen
  \bibfield  {author} {\bibinfo {author} {\bibfnamefont {C.}~\bibnamefont
  {Cheung}}, \bibinfo {author} {\bibfnamefont {P.}~\bibnamefont {Creminelli}},
  \bibinfo {author} {\bibfnamefont {A.~L.}\ \bibnamefont {Fitzpatrick}},
  \bibinfo {author} {\bibfnamefont {J.}~\bibnamefont {Kaplan}}, \ and\ \bibinfo
  {author} {\bibfnamefont {L.}~\bibnamefont {Senatore}},\ }\href {\doibase
  10.1088/1126-6708/2008/03/014} {\bibfield  {journal} {\bibinfo  {journal}
  {JHEP}\ }\textbf {\bibinfo {volume} {03}},\ \bibinfo {pages} {014} (\bibinfo
  {year} {2008})},\ \Eprint {http://arxiv.org/abs/0709.0293} {arXiv:0709.0293
  [hep-th]} \BibitemShut {NoStop}%
%%CITATION = ARXIV:0709.0293;%%
\bibitem [{\citenamefont {Gleyzes}\ \emph {et~al.}(2013)\citenamefont
  {Gleyzes}, \citenamefont {Langlois}, \citenamefont {Piazza},\ and\
  \citenamefont {Vernizzi}}]{Gleyzes:2013ooa}%
  \BibitemOpen
  \bibfield  {author} {\bibinfo {author} {\bibfnamefont {J.}~\bibnamefont
  {Gleyzes}}, \bibinfo {author} {\bibfnamefont {D.}~\bibnamefont {Langlois}},
  \bibinfo {author} {\bibfnamefont {F.}~\bibnamefont {Piazza}}, \ and\ \bibinfo
  {author} {\bibfnamefont {F.}~\bibnamefont {Vernizzi}},\ }\href {\doibase
  10.1088/1475-7516/2013/08/025} {\bibfield  {journal} {\bibinfo  {journal}
  {JCAP}\ }\textbf {\bibinfo {volume} {1308}},\ \bibinfo {pages} {025}
  (\bibinfo {year} {2013})},\ \Eprint {http://arxiv.org/abs/1304.4840}
  {arXiv:1304.4840 [hep-th]} \BibitemShut {NoStop}%
%%CITATION = ARXIV:1304.4840;%%
\bibitem [{\citenamefont {Kase}\ and\ \citenamefont
  {Tsujikawa}(2014)}]{Kase:2014cwa}%
  \BibitemOpen
  \bibfield  {author} {\bibinfo {author} {\bibfnamefont {R.}~\bibnamefont
  {Kase}}\ and\ \bibinfo {author} {\bibfnamefont {S.}~\bibnamefont
  {Tsujikawa}},\ }\href {\doibase 10.1142/S0218271814430081} {\bibfield
  {journal} {\bibinfo  {journal} {Int. J. Mod. Phys.}\ }\textbf {\bibinfo
  {volume} {D23}},\ \bibinfo {pages} {1443008} (\bibinfo {year} {2014})},\
  \Eprint {http://arxiv.org/abs/1409.1984} {arXiv:1409.1984 [hep-th]}
  \BibitemShut {NoStop}%
%%CITATION = ARXIV:1409.1984;%%
\bibitem [{\citenamefont {Gleyzes}\ \emph {et~al.}(2014)\citenamefont
  {Gleyzes}, \citenamefont {Langlois},\ and\ \citenamefont
  {Vernizzi}}]{Gleyzes:2014rba}%
  \BibitemOpen
  \bibfield  {author} {\bibinfo {author} {\bibfnamefont {J.}~\bibnamefont
  {Gleyzes}}, \bibinfo {author} {\bibfnamefont {D.}~\bibnamefont {Langlois}}, \
  and\ \bibinfo {author} {\bibfnamefont {F.}~\bibnamefont {Vernizzi}},\ }\href
  {\doibase 10.1142/S021827181443010X} {\bibfield  {journal} {\bibinfo
  {journal} {Int. J. Mod. Phys.}\ }\textbf {\bibinfo {volume} {D23}},\ \bibinfo
  {pages} {1443010} (\bibinfo {year} {2014})},\ \Eprint
  {http://arxiv.org/abs/1411.3712} {arXiv:1411.3712 [hep-th]} \BibitemShut
  {NoStop}%
%%CITATION = ARXIV:1411.3712;%%
\bibitem [{\citenamefont {Gleyzes}\ \emph
  {et~al.}(2015{\natexlab{a}})\citenamefont {Gleyzes}, \citenamefont
  {Langlois}, \citenamefont {Mancarella},\ and\ \citenamefont
  {Vernizzi}}]{Gleyzes:2015pma}%
  \BibitemOpen
  \bibfield  {author} {\bibinfo {author} {\bibfnamefont {J.}~\bibnamefont
  {Gleyzes}}, \bibinfo {author} {\bibfnamefont {D.}~\bibnamefont {Langlois}},
  \bibinfo {author} {\bibfnamefont {M.}~\bibnamefont {Mancarella}}, \ and\
  \bibinfo {author} {\bibfnamefont {F.}~\bibnamefont {Vernizzi}},\ }\href
  {\doibase 10.1088/1475-7516/2015/08/054} {\bibfield  {journal} {\bibinfo
  {journal} {JCAP}\ }\textbf {\bibinfo {volume} {1508}},\ \bibinfo {pages}
  {054} (\bibinfo {year} {2015}{\natexlab{a}})},\ \Eprint
  {http://arxiv.org/abs/1504.05481} {arXiv:1504.05481 [astro-ph.CO]}
  \BibitemShut {NoStop}%
%%CITATION = ARXIV:1504.05481;%%
\bibitem [{\citenamefont {Horndeski}(1974)}]{Horndeski:1974wa}%
  \BibitemOpen
  \bibfield  {author} {\bibinfo {author} {\bibfnamefont {G.~W.}\ \bibnamefont
  {Horndeski}},\ }\href {\doibase 10.1007/BF01807638} {\bibfield  {journal}
  {\bibinfo  {journal} {Int. J. Theor. Phys.}\ }\textbf {\bibinfo {volume}
  {10}},\ \bibinfo {pages} {363} (\bibinfo {year} {1974})}\BibitemShut
  {NoStop}%
%%CITATION = IJTPB,10,363;%%
\bibitem [{\citenamefont {Nicolis}\ \emph {et~al.}(2009)\citenamefont
  {Nicolis}, \citenamefont {Rattazzi},\ and\ \citenamefont
  {Trincherini}}]{Nicolis:2008in}%
  \BibitemOpen
  \bibfield  {author} {\bibinfo {author} {\bibfnamefont {A.}~\bibnamefont
  {Nicolis}}, \bibinfo {author} {\bibfnamefont {R.}~\bibnamefont {Rattazzi}}, \
  and\ \bibinfo {author} {\bibfnamefont {E.}~\bibnamefont {Trincherini}},\
  }\href {\doibase 10.1103/PhysRevD.79.064036} {\bibfield  {journal} {\bibinfo
  {journal} {Phys. Rev.}\ }\textbf {\bibinfo {volume} {D79}},\ \bibinfo {pages}
  {064036} (\bibinfo {year} {2009})},\ \Eprint {http://arxiv.org/abs/0811.2197}
  {arXiv:0811.2197 [hep-th]} \BibitemShut {NoStop}%
%%CITATION = ARXIV:0811.2197;%%
\bibitem [{\citenamefont {Deffayet}\ \emph
  {et~al.}(2009{\natexlab{a}})\citenamefont {Deffayet}, \citenamefont
  {Esposito-Farese},\ and\ \citenamefont {Vikman}}]{Deffayet:2009wt}%
  \BibitemOpen
  \bibfield  {author} {\bibinfo {author} {\bibfnamefont {C.}~\bibnamefont
  {Deffayet}}, \bibinfo {author} {\bibfnamefont {G.}~\bibnamefont
  {Esposito-Farese}}, \ and\ \bibinfo {author} {\bibfnamefont {A.}~\bibnamefont
  {Vikman}},\ }\href {\doibase 10.1103/PhysRevD.79.084003} {\bibfield
  {journal} {\bibinfo  {journal} {Phys. Rev.}\ }\textbf {\bibinfo {volume}
  {D79}},\ \bibinfo {pages} {084003} (\bibinfo {year} {2009}{\natexlab{a}})},\
  \Eprint {http://arxiv.org/abs/0901.1314} {arXiv:0901.1314 [hep-th]}
  \BibitemShut {NoStop}%
%%CITATION = ARXIV:0901.1314;%%
\bibitem [{\citenamefont {Deffayet}\ \emph
  {et~al.}(2009{\natexlab{b}})\citenamefont {Deffayet}, \citenamefont {Deser},\
  and\ \citenamefont {Esposito-Farese}}]{Deffayet:2009mn}%
  \BibitemOpen
  \bibfield  {author} {\bibinfo {author} {\bibfnamefont {C.}~\bibnamefont
  {Deffayet}}, \bibinfo {author} {\bibfnamefont {S.}~\bibnamefont {Deser}}, \
  and\ \bibinfo {author} {\bibfnamefont {G.}~\bibnamefont {Esposito-Farese}},\
  }\href {\doibase 10.1103/PhysRevD.80.064015} {\bibfield  {journal} {\bibinfo
  {journal} {Phys. Rev.}\ }\textbf {\bibinfo {volume} {D80}},\ \bibinfo {pages}
  {064015} (\bibinfo {year} {2009}{\natexlab{b}})},\ \Eprint
  {http://arxiv.org/abs/0906.1967} {arXiv:0906.1967 [gr-qc]} \BibitemShut
  {NoStop}%
%%CITATION = ARXIV:0906.1967;%%
\bibitem [{\citenamefont {Deffayet}\ \emph {et~al.}(2011)\citenamefont
  {Deffayet}, \citenamefont {Gao}, \citenamefont {Steer},\ and\ \citenamefont
  {Zahariade}}]{Deffayet:2011gz}%
  \BibitemOpen
  \bibfield  {author} {\bibinfo {author} {\bibfnamefont {C.}~\bibnamefont
  {Deffayet}}, \bibinfo {author} {\bibfnamefont {X.}~\bibnamefont {Gao}},
  \bibinfo {author} {\bibfnamefont {D.~A.}\ \bibnamefont {Steer}}, \ and\
  \bibinfo {author} {\bibfnamefont {G.}~\bibnamefont {Zahariade}},\ }\href
  {\doibase 10.1103/PhysRevD.84.064039} {\bibfield  {journal} {\bibinfo
  {journal} {Phys. Rev.}\ }\textbf {\bibinfo {volume} {D84}},\ \bibinfo {pages}
  {064039} (\bibinfo {year} {2011})},\ \Eprint {http://arxiv.org/abs/1103.3260}
  {arXiv:1103.3260 [hep-th]} \BibitemShut {NoStop}%
%%CITATION = ARXIV:1103.3260;%%
\bibitem [{\citenamefont {Kobayashi}\ \emph {et~al.}(2011)\citenamefont
  {Kobayashi}, \citenamefont {Yamaguchi},\ and\ \citenamefont
  {Yokoyama}}]{Kobayashi:2011nu}%
  \BibitemOpen
  \bibfield  {author} {\bibinfo {author} {\bibfnamefont {T.}~\bibnamefont
  {Kobayashi}}, \bibinfo {author} {\bibfnamefont {M.}~\bibnamefont
  {Yamaguchi}}, \ and\ \bibinfo {author} {\bibfnamefont {J.}~\bibnamefont
  {Yokoyama}},\ }\href {\doibase 10.1143/PTP.126.511} {\bibfield  {journal}
  {\bibinfo  {journal} {Prog. Theor. Phys.}\ }\textbf {\bibinfo {volume}
  {126}},\ \bibinfo {pages} {511} (\bibinfo {year} {2011})},\ \Eprint
  {http://arxiv.org/abs/1105.5723} {arXiv:1105.5723 [hep-th]} \BibitemShut
  {NoStop}%
%%CITATION = ARXIV:1105.5723;%%
\bibitem [{\citenamefont {Gleyzes}\ \emph
  {et~al.}(2015{\natexlab{b}})\citenamefont {Gleyzes}, \citenamefont
  {Langlois}, \citenamefont {Piazza},\ and\ \citenamefont
  {Vernizzi}}]{Gleyzes:2014dya}%
  \BibitemOpen
  \bibfield  {author} {\bibinfo {author} {\bibfnamefont {J.}~\bibnamefont
  {Gleyzes}}, \bibinfo {author} {\bibfnamefont {D.}~\bibnamefont {Langlois}},
  \bibinfo {author} {\bibfnamefont {F.}~\bibnamefont {Piazza}}, \ and\ \bibinfo
  {author} {\bibfnamefont {F.}~\bibnamefont {Vernizzi}},\ }\href {\doibase
  10.1103/PhysRevLett.114.211101} {\bibfield  {journal} {\bibinfo  {journal}
  {Phys. Rev. Lett.}\ }\textbf {\bibinfo {volume} {114}},\ \bibinfo {pages}
  {211101} (\bibinfo {year} {2015}{\natexlab{b}})},\ \Eprint
  {http://arxiv.org/abs/1404.6495} {arXiv:1404.6495 [hep-th]} \BibitemShut
  {NoStop}%
%%CITATION = ARXIV:1404.6495;%%
\bibitem [{\citenamefont {Gleyzes}\ \emph
  {et~al.}(2015{\natexlab{c}})\citenamefont {Gleyzes}, \citenamefont
  {Langlois}, \citenamefont {Piazza},\ and\ \citenamefont
  {Vernizzi}}]{Gleyzes:2014qga}%
  \BibitemOpen
  \bibfield  {author} {\bibinfo {author} {\bibfnamefont {J.}~\bibnamefont
  {Gleyzes}}, \bibinfo {author} {\bibfnamefont {D.}~\bibnamefont {Langlois}},
  \bibinfo {author} {\bibfnamefont {F.}~\bibnamefont {Piazza}}, \ and\ \bibinfo
  {author} {\bibfnamefont {F.}~\bibnamefont {Vernizzi}},\ }\href {\doibase
  10.1088/1475-7516/2015/02/018} {\bibfield  {journal} {\bibinfo  {journal}
  {JCAP}\ }\textbf {\bibinfo {volume} {1502}},\ \bibinfo {pages} {018}
  (\bibinfo {year} {2015}{\natexlab{c}})},\ \Eprint
  {http://arxiv.org/abs/1408.1952} {arXiv:1408.1952 [astro-ph.CO]} \BibitemShut
  {NoStop}%
%%CITATION = ARXIV:1408.1952;%%
\bibitem [{\citenamefont {Horava}(2009)}]{Horava:2009uw}%
  \BibitemOpen
  \bibfield  {author} {\bibinfo {author} {\bibfnamefont {P.}~\bibnamefont
  {Horava}},\ }\href {\doibase 10.1103/PhysRevD.79.084008} {\bibfield
  {journal} {\bibinfo  {journal} {Phys. Rev.}\ }\textbf {\bibinfo {volume}
  {D79}},\ \bibinfo {pages} {084008} (\bibinfo {year} {2009})},\ \Eprint
  {http://arxiv.org/abs/0901.3775} {arXiv:0901.3775 [hep-th]} \BibitemShut
  {NoStop}%
%%CITATION = ARXIV:0901.3775;%%
\bibitem [{\citenamefont {Blas}\ \emph
  {et~al.}(2010{\natexlab{a}})\citenamefont {Blas}, \citenamefont {Pujolas},\
  and\ \citenamefont {Sibiryakov}}]{Blas:2009qj}%
  \BibitemOpen
  \bibfield  {author} {\bibinfo {author} {\bibfnamefont {D.}~\bibnamefont
  {Blas}}, \bibinfo {author} {\bibfnamefont {O.}~\bibnamefont {Pujolas}}, \
  and\ \bibinfo {author} {\bibfnamefont {S.}~\bibnamefont {Sibiryakov}},\
  }\href {\doibase 10.1103/PhysRevLett.104.181302} {\bibfield  {journal}
  {\bibinfo  {journal} {Phys. Rev. Lett.}\ }\textbf {\bibinfo {volume} {104}},\
  \bibinfo {pages} {181302} (\bibinfo {year} {2010}{\natexlab{a}})},\ \Eprint
  {http://arxiv.org/abs/0909.3525} {arXiv:0909.3525 [hep-th]} \BibitemShut
  {NoStop}%
%%CITATION = ARXIV:0909.3525;%%
\bibitem [{\citenamefont {Blas}\ \emph
  {et~al.}(2010{\natexlab{b}})\citenamefont {Blas}, \citenamefont {Pujolas},\
  and\ \citenamefont {Sibiryakov}}]{Blas:2009ck}%
  \BibitemOpen
  \bibfield  {author} {\bibinfo {author} {\bibfnamefont {D.}~\bibnamefont
  {Blas}}, \bibinfo {author} {\bibfnamefont {O.}~\bibnamefont {Pujolas}}, \
  and\ \bibinfo {author} {\bibfnamefont {S.}~\bibnamefont {Sibiryakov}},\
  }\href {\doibase 10.1016/j.physletb.2010.03.073} {\bibfield  {journal}
  {\bibinfo  {journal} {Phys. Lett.}\ }\textbf {\bibinfo {volume} {B688}},\
  \bibinfo {pages} {350} (\bibinfo {year} {2010}{\natexlab{b}})},\ \Eprint
  {http://arxiv.org/abs/0912.0550} {arXiv:0912.0550 [hep-th]} \BibitemShut
  {NoStop}%
%%CITATION = ARXIV:0912.0550;%%
\bibitem [{\citenamefont {Langlois}\ and\ \citenamefont
  {Noui}(2016)}]{Langlois:2015cwa}%
  \BibitemOpen
  \bibfield  {author} {\bibinfo {author} {\bibfnamefont {D.}~\bibnamefont
  {Langlois}}\ and\ \bibinfo {author} {\bibfnamefont {K.}~\bibnamefont
  {Noui}},\ }\href {\doibase 10.1088/1475-7516/2016/02/034} {\bibfield
  {journal} {\bibinfo  {journal} {JCAP}\ }\textbf {\bibinfo {volume} {1602}},\
  \bibinfo {pages} {034} (\bibinfo {year} {2016})},\ \Eprint
  {http://arxiv.org/abs/1510.06930} {arXiv:1510.06930 [gr-qc]} \BibitemShut
  {NoStop}%
%%CITATION = ARXIV:1510.06930;%%
\bibitem [{\citenamefont {Crisostomi}\ \emph {et~al.}(2016)\citenamefont
  {Crisostomi}, \citenamefont {Koyama},\ and\ \citenamefont
  {Tasinato}}]{Crisostomi:2016czh}%
  \BibitemOpen
  \bibfield  {author} {\bibinfo {author} {\bibfnamefont {M.}~\bibnamefont
  {Crisostomi}}, \bibinfo {author} {\bibfnamefont {K.}~\bibnamefont {Koyama}},
  \ and\ \bibinfo {author} {\bibfnamefont {G.}~\bibnamefont {Tasinato}},\
  }\href {\doibase 10.1088/1475-7516/2016/04/044} {\bibfield  {journal}
  {\bibinfo  {journal} {JCAP}\ }\textbf {\bibinfo {volume} {1604}},\ \bibinfo
  {pages} {044} (\bibinfo {year} {2016})},\ \Eprint
  {http://arxiv.org/abs/1602.03119} {arXiv:1602.03119 [hep-th]} \BibitemShut
  {NoStop}%
%%CITATION = ARXIV:1602.03119;%%
\bibitem [{\citenamefont {Ben~Achour}\ \emph
  {et~al.}(2016{\natexlab{a}})\citenamefont {Ben~Achour}, \citenamefont
  {Langlois},\ and\ \citenamefont {Noui}}]{Achour:2016rkg}%
  \BibitemOpen
  \bibfield  {author} {\bibinfo {author} {\bibfnamefont {J.}~\bibnamefont
  {Ben~Achour}}, \bibinfo {author} {\bibfnamefont {D.}~\bibnamefont
  {Langlois}}, \ and\ \bibinfo {author} {\bibfnamefont {K.}~\bibnamefont
  {Noui}},\ }\href {\doibase 10.1103/PhysRevD.93.124005} {\bibfield  {journal}
  {\bibinfo  {journal} {Phys. Rev.}\ }\textbf {\bibinfo {volume} {D93}},\
  \bibinfo {pages} {124005} (\bibinfo {year} {2016}{\natexlab{a}})},\ \Eprint
  {http://arxiv.org/abs/1602.08398} {arXiv:1602.08398 [gr-qc]} \BibitemShut
  {NoStop}%
%%CITATION = ARXIV:1602.08398;%%
\bibitem [{\citenamefont {Ben~Achour}\ \emph
  {et~al.}(2016{\natexlab{b}})\citenamefont {Ben~Achour}, \citenamefont
  {Crisostomi}, \citenamefont {Koyama}, \citenamefont {Langlois}, \citenamefont
  {Noui},\ and\ \citenamefont {Tasinato}}]{BenAchour:2016fzp}%
  \BibitemOpen
  \bibfield  {author} {\bibinfo {author} {\bibfnamefont {J.}~\bibnamefont
  {Ben~Achour}}, \bibinfo {author} {\bibfnamefont {M.}~\bibnamefont
  {Crisostomi}}, \bibinfo {author} {\bibfnamefont {K.}~\bibnamefont {Koyama}},
  \bibinfo {author} {\bibfnamefont {D.}~\bibnamefont {Langlois}}, \bibinfo
  {author} {\bibfnamefont {K.}~\bibnamefont {Noui}}, \ and\ \bibinfo {author}
  {\bibfnamefont {G.}~\bibnamefont {Tasinato}},\ }\href {\doibase
  10.1007/JHEP12(2016)100} {\bibfield  {journal} {\bibinfo  {journal} {JHEP}\
  }\textbf {\bibinfo {volume} {12}},\ \bibinfo {pages} {100} (\bibinfo {year}
  {2016}{\natexlab{b}})},\ \Eprint {http://arxiv.org/abs/1608.08135}
  {arXiv:1608.08135 [hep-th]} \BibitemShut {NoStop}%
%%CITATION = ARXIV:1608.08135;%%
\bibitem [{\citenamefont {Langlois}\ \emph {et~al.}(2017)\citenamefont
  {Langlois}, \citenamefont {Mancarella}, \citenamefont {Noui},\ and\
  \citenamefont {Vernizzi}}]{Langlois:2017mxy}%
  \BibitemOpen
  \bibfield  {author} {\bibinfo {author} {\bibfnamefont {D.}~\bibnamefont
  {Langlois}}, \bibinfo {author} {\bibfnamefont {M.}~\bibnamefont
  {Mancarella}}, \bibinfo {author} {\bibfnamefont {K.}~\bibnamefont {Noui}}, \
  and\ \bibinfo {author} {\bibfnamefont {F.}~\bibnamefont {Vernizzi}},\ }\href
  {\doibase 10.1088/1475-7516/2017/05/033} {\bibfield  {journal} {\bibinfo
  {journal} {JCAP}\ }\textbf {\bibinfo {volume} {1705}},\ \bibinfo {pages}
  {033} (\bibinfo {year} {2017})},\ \Eprint {http://arxiv.org/abs/1703.03797}
  {arXiv:1703.03797 [hep-th]} \BibitemShut {NoStop}%
%%CITATION = ARXIV:1703.03797;%%
\bibitem [{\citenamefont {Motohashi}\ and\ \citenamefont
  {Suyama}(2015)}]{Motohashi:2014opa}%
  \BibitemOpen
  \bibfield  {author} {\bibinfo {author} {\bibfnamefont {H.}~\bibnamefont
  {Motohashi}}\ and\ \bibinfo {author} {\bibfnamefont {T.}~\bibnamefont
  {Suyama}},\ }\href {\doibase 10.1103/PhysRevD.91.085009} {\bibfield
  {journal} {\bibinfo  {journal} {Phys. Rev.}\ }\textbf {\bibinfo {volume}
  {D91}},\ \bibinfo {pages} {085009} (\bibinfo {year} {2015})},\ \Eprint
  {http://arxiv.org/abs/1411.3721} {arXiv:1411.3721 [physics.class-ph]}
  \BibitemShut {NoStop}%
%%CITATION = ARXIV:1411.3721;%%
\bibitem [{\citenamefont {Motohashi}\ \emph
  {et~al.}(2016{\natexlab{a}})\citenamefont {Motohashi}, \citenamefont {Noui},
  \citenamefont {Suyama}, \citenamefont {Yamaguchi},\ and\ \citenamefont
  {Langlois}}]{Motohashi:2016ftl}%
  \BibitemOpen
  \bibfield  {author} {\bibinfo {author} {\bibfnamefont {H.}~\bibnamefont
  {Motohashi}}, \bibinfo {author} {\bibfnamefont {K.}~\bibnamefont {Noui}},
  \bibinfo {author} {\bibfnamefont {T.}~\bibnamefont {Suyama}}, \bibinfo
  {author} {\bibfnamefont {M.}~\bibnamefont {Yamaguchi}}, \ and\ \bibinfo
  {author} {\bibfnamefont {D.}~\bibnamefont {Langlois}},\ }\href {\doibase
  10.1088/1475-7516/2016/07/033} {\bibfield  {journal} {\bibinfo  {journal}
  {JCAP}\ }\textbf {\bibinfo {volume} {1607}},\ \bibinfo {pages} {033}
  (\bibinfo {year} {2016}{\natexlab{a}})},\ \Eprint
  {http://arxiv.org/abs/1603.09355} {arXiv:1603.09355 [hep-th]} \BibitemShut
  {NoStop}%
%%CITATION = ARXIV:1603.09355;%%
\bibitem [{\citenamefont {Klein}\ and\ \citenamefont
  {Roest}(2016)}]{Klein:2016aiq}%
  \BibitemOpen
  \bibfield  {author} {\bibinfo {author} {\bibfnamefont {R.}~\bibnamefont
  {Klein}}\ and\ \bibinfo {author} {\bibfnamefont {D.}~\bibnamefont {Roest}},\
  }\href {\doibase 10.1007/JHEP07(2016)130} {\bibfield  {journal} {\bibinfo
  {journal} {JHEP}\ }\textbf {\bibinfo {volume} {07}},\ \bibinfo {pages} {130}
  (\bibinfo {year} {2016})},\ \Eprint {http://arxiv.org/abs/1604.01719}
  {arXiv:1604.01719 [hep-th]} \BibitemShut {NoStop}%
%%CITATION = ARXIV:1604.01719;%%
\bibitem [{\citenamefont {Crisostomi}\ \emph {et~al.}(2017)\citenamefont
  {Crisostomi}, \citenamefont {Klein},\ and\ \citenamefont
  {Roest}}]{Crisostomi:2017aim}%
  \BibitemOpen
  \bibfield  {author} {\bibinfo {author} {\bibfnamefont {M.}~\bibnamefont
  {Crisostomi}}, \bibinfo {author} {\bibfnamefont {R.}~\bibnamefont {Klein}}, \
  and\ \bibinfo {author} {\bibfnamefont {D.}~\bibnamefont {Roest}},\ }\href
  {\doibase 10.1007/JHEP06(2017)124} {\bibfield  {journal} {\bibinfo  {journal}
  {JHEP}\ }\textbf {\bibinfo {volume} {06}},\ \bibinfo {pages} {124} (\bibinfo
  {year} {2017})},\ \Eprint {http://arxiv.org/abs/1703.01623} {arXiv:1703.01623
  [hep-th]} \BibitemShut {NoStop}%
%%CITATION = ARXIV:1703.01623;%%
\bibitem [{\citenamefont {Gao}(2014{\natexlab{a}})}]{Gao:2014soa}%
  \BibitemOpen
  \bibfield  {author} {\bibinfo {author} {\bibfnamefont {X.}~\bibnamefont
  {Gao}},\ }\href {\doibase 10.1103/PhysRevD.90.081501} {\bibfield  {journal}
  {\bibinfo  {journal} {Phys. Rev.}\ }\textbf {\bibinfo {volume} {D90}},\
  \bibinfo {pages} {081501} (\bibinfo {year} {2014}{\natexlab{a}})},\ \Eprint
  {http://arxiv.org/abs/1406.0822} {arXiv:1406.0822 [gr-qc]} \BibitemShut
  {NoStop}%
%%CITATION = ARXIV:1406.0822;%%
\bibitem [{\citenamefont {Gao}(2014{\natexlab{b}})}]{Gao:2014fra}%
  \BibitemOpen
  \bibfield  {author} {\bibinfo {author} {\bibfnamefont {X.}~\bibnamefont
  {Gao}},\ }\href {\doibase 10.1103/PhysRevD.90.104033} {\bibfield  {journal}
  {\bibinfo  {journal} {Phys. Rev.}\ }\textbf {\bibinfo {volume} {D90}},\
  \bibinfo {pages} {104033} (\bibinfo {year} {2014}{\natexlab{b}})},\ \Eprint
  {http://arxiv.org/abs/1409.6708} {arXiv:1409.6708 [gr-qc]} \BibitemShut
  {NoStop}%
%%CITATION = ARXIV:1409.6708;%%
\bibitem [{\citenamefont {Stewart}(2002)}]{Stewart:2001cd}%
  \BibitemOpen
  \bibfield  {author} {\bibinfo {author} {\bibfnamefont {E.~D.}\ \bibnamefont
  {Stewart}},\ }\href {\doibase 10.1103/PhysRevD.65.103508} {\bibfield
  {journal} {\bibinfo  {journal} {Phys.Rev.}\ }\textbf {\bibinfo {volume}
  {D65}},\ \bibinfo {pages} {103508} (\bibinfo {year} {2002})},\ \Eprint
  {http://arxiv.org/abs/astro-ph/0110322} {arXiv:astro-ph/0110322 [astro-ph]}
  \BibitemShut {NoStop}%
%%CITATION = ASTRO-PH/0110322;%%
\bibitem [{\citenamefont {Gong}(2004)}]{Gong:2004kd}%
  \BibitemOpen
  \bibfield  {author} {\bibinfo {author} {\bibfnamefont {J.-O.}\ \bibnamefont
  {Gong}},\ }\href {\doibase 10.1088/0264-9381/21/23/016} {\bibfield  {journal}
  {\bibinfo  {journal} {Class. Quant. Grav.}\ }\textbf {\bibinfo {volume}
  {21}},\ \bibinfo {pages} {5555} (\bibinfo {year} {2004})},\ \Eprint
  {http://arxiv.org/abs/gr-qc/0408039} {arXiv:gr-qc/0408039 [gr-qc]}
  \BibitemShut {NoStop}%
%%CITATION = GR-QC/0408039;%%
\bibitem [{\citenamefont {Hu}(2011)}]{Hu:2011vr}%
  \BibitemOpen
  \bibfield  {author} {\bibinfo {author} {\bibfnamefont {W.}~\bibnamefont
  {Hu}},\ }\href {\doibase 10.1103/PhysRevD.84.027303} {\bibfield  {journal}
  {\bibinfo  {journal} {Phys. Rev.}\ }\textbf {\bibinfo {volume} {D84}},\
  \bibinfo {pages} {027303} (\bibinfo {year} {2011})},\ \Eprint
  {http://arxiv.org/abs/1104.4500} {arXiv:1104.4500 [astro-ph.CO]} \BibitemShut
  {NoStop}%
%%CITATION = ARXIV:1104.4500;%%
\bibitem [{\citenamefont {Hu}(2014)}]{Hu:2014hoa}%
  \BibitemOpen
  \bibfield  {author} {\bibinfo {author} {\bibfnamefont {W.}~\bibnamefont
  {Hu}},\ }\href {\doibase 10.1103/PhysRevD.89.123503} {\bibfield  {journal}
  {\bibinfo  {journal} {Phys. Rev.}\ }\textbf {\bibinfo {volume} {D89}},\
  \bibinfo {pages} {123503} (\bibinfo {year} {2014})},\ \Eprint
  {http://arxiv.org/abs/1405.2020} {arXiv:1405.2020 [astro-ph.CO]} \BibitemShut
  {NoStop}%
%%CITATION = ARXIV:1405.2020;%%
\bibitem [{\citenamefont {Dvorkin}\ and\ \citenamefont
  {Hu}(2011)}]{Dvorkin:2011ui}%
  \BibitemOpen
  \bibfield  {author} {\bibinfo {author} {\bibfnamefont {C.}~\bibnamefont
  {Dvorkin}}\ and\ \bibinfo {author} {\bibfnamefont {W.}~\bibnamefont {Hu}},\
  }\href {\doibase 10.1103/PhysRevD.84.063515} {\bibfield  {journal} {\bibinfo
  {journal} {Phys. Rev.}\ }\textbf {\bibinfo {volume} {D84}},\ \bibinfo {pages}
  {063515} (\bibinfo {year} {2011})},\ \Eprint {http://arxiv.org/abs/1106.4016}
  {arXiv:1106.4016 [astro-ph.CO]} \BibitemShut {NoStop}%
%%CITATION = ARXIV:1106.4016;%%
\bibitem [{\citenamefont {Miranda}\ \emph {et~al.}(2015)\citenamefont
  {Miranda}, \citenamefont {Hu},\ and\ \citenamefont
  {Dvorkin}}]{Miranda:2014fwa}%
  \BibitemOpen
  \bibfield  {author} {\bibinfo {author} {\bibfnamefont {V.}~\bibnamefont
  {Miranda}}, \bibinfo {author} {\bibfnamefont {W.}~\bibnamefont {Hu}}, \ and\
  \bibinfo {author} {\bibfnamefont {C.}~\bibnamefont {Dvorkin}},\ }\href
  {\doibase 10.1103/PhysRevD.91.063514} {\bibfield  {journal} {\bibinfo
  {journal} {Phys. Rev.}\ }\textbf {\bibinfo {volume} {D91}},\ \bibinfo {pages}
  {063514} (\bibinfo {year} {2015})},\ \Eprint {http://arxiv.org/abs/1411.5956}
  {arXiv:1411.5956 [astro-ph.CO]} \BibitemShut {NoStop}%
%%CITATION = ARXIV:1411.5956;%%
\bibitem [{\citenamefont {Gubitosi}\ \emph {et~al.}(2013)\citenamefont
  {Gubitosi}, \citenamefont {Piazza},\ and\ \citenamefont
  {Vernizzi}}]{Gubitosi:2012hu}%
  \BibitemOpen
  \bibfield  {author} {\bibinfo {author} {\bibfnamefont {G.}~\bibnamefont
  {Gubitosi}}, \bibinfo {author} {\bibfnamefont {F.}~\bibnamefont {Piazza}}, \
  and\ \bibinfo {author} {\bibfnamefont {F.}~\bibnamefont {Vernizzi}},\ }\href
  {\doibase 10.1088/1475-7516/2013/02/032} {\bibfield  {journal} {\bibinfo
  {journal} {JCAP}\ }\textbf {\bibinfo {volume} {1302}},\ \bibinfo {pages}
  {032} (\bibinfo {year} {2013})},\ \Eprint {http://arxiv.org/abs/1210.0201}
  {arXiv:1210.0201 [hep-th]} \BibitemShut {NoStop}%
%%CITATION = ARXIV:1210.0201;%%
\bibitem [{\citenamefont {Motohashi}\ \emph
  {et~al.}(2016{\natexlab{b}})\citenamefont {Motohashi}, \citenamefont
  {Suyama},\ and\ \citenamefont {Takahashi}}]{Motohashi:2016prk}%
  \BibitemOpen
  \bibfield  {author} {\bibinfo {author} {\bibfnamefont {H.}~\bibnamefont
  {Motohashi}}, \bibinfo {author} {\bibfnamefont {T.}~\bibnamefont {Suyama}}, \
  and\ \bibinfo {author} {\bibfnamefont {K.}~\bibnamefont {Takahashi}},\ }\href
  {\doibase 10.1103/PhysRevD.94.124021} {\bibfield  {journal} {\bibinfo
  {journal} {Phys. Rev.}\ }\textbf {\bibinfo {volume} {D94}},\ \bibinfo {pages}
  {124021} (\bibinfo {year} {2016}{\natexlab{b}})},\ \Eprint
  {http://arxiv.org/abs/1608.00071} {arXiv:1608.00071 [gr-qc]} \BibitemShut
  {NoStop}%
%%CITATION = ARXIV:1608.00071;%%
\bibitem [{\citenamefont {Alishahiha}\ \emph {et~al.}(2011)\citenamefont
  {Alishahiha}, \citenamefont {Firouzjahi},\ and\ \citenamefont
  {Namjoo}}]{Alishahiha:2011yh}%
  \BibitemOpen
  \bibfield  {author} {\bibinfo {author} {\bibfnamefont {M.}~\bibnamefont
  {Alishahiha}}, \bibinfo {author} {\bibfnamefont {H.}~\bibnamefont
  {Firouzjahi}}, \ and\ \bibinfo {author} {\bibfnamefont {M.~H.}\ \bibnamefont
  {Namjoo}},\ }\href {\doibase 10.1088/1475-7516/2011/08/028} {\bibfield
  {journal} {\bibinfo  {journal} {JCAP}\ }\textbf {\bibinfo {volume} {1108}},\
  \bibinfo {pages} {028} (\bibinfo {year} {2011})},\ \Eprint
  {http://arxiv.org/abs/1103.2919} {arXiv:1103.2919 [hep-th]} \BibitemShut
  {NoStop}%
%%CITATION = ARXIV:1103.2919;%%
\bibitem [{\citenamefont {Alishahiha}\ \emph {et~al.}(2013)\citenamefont
  {Alishahiha}, \citenamefont {Firouzjahi}, \citenamefont {Koyama},\ and\
  \citenamefont {Namjoo}}]{Alishahiha:2013nsa}%
  \BibitemOpen
  \bibfield  {author} {\bibinfo {author} {\bibfnamefont {M.}~\bibnamefont
  {Alishahiha}}, \bibinfo {author} {\bibfnamefont {H.}~\bibnamefont
  {Firouzjahi}}, \bibinfo {author} {\bibfnamefont {K.}~\bibnamefont {Koyama}},
  \ and\ \bibinfo {author} {\bibfnamefont {M.~H.}\ \bibnamefont {Namjoo}},\
  }\href {\doibase 10.1103/PhysRevD.88.103512} {\bibfield  {journal} {\bibinfo
  {journal} {Phys. Rev.}\ }\textbf {\bibinfo {volume} {D88}},\ \bibinfo {pages}
  {103512} (\bibinfo {year} {2013})},\ \Eprint {http://arxiv.org/abs/1305.4327}
  {arXiv:1305.4327 [hep-th]} \BibitemShut {NoStop}%
%%CITATION = ARXIV:1305.4327;%%
\bibitem [{\citenamefont {Kinney}(2005)}]{Kinney:2005vj}%
  \BibitemOpen
  \bibfield  {author} {\bibinfo {author} {\bibfnamefont {W.~H.}\ \bibnamefont
  {Kinney}},\ }\href {\doibase 10.1103/PhysRevD.72.023515} {\bibfield
  {journal} {\bibinfo  {journal} {Phys. Rev.}\ }\textbf {\bibinfo {volume}
  {D72}},\ \bibinfo {pages} {023515} (\bibinfo {year} {2005})},\ \Eprint
  {http://arxiv.org/abs/gr-qc/0503017} {arXiv:gr-qc/0503017 [gr-qc]}
  \BibitemShut {NoStop}%
%%CITATION = GR-QC/0503017;%%
\bibitem [{\citenamefont {Namjoo}\ \emph {et~al.}(2013)\citenamefont {Namjoo},
  \citenamefont {Firouzjahi},\ and\ \citenamefont {Sasaki}}]{Namjoo:2012aa}%
  \BibitemOpen
  \bibfield  {author} {\bibinfo {author} {\bibfnamefont {M.~H.}\ \bibnamefont
  {Namjoo}}, \bibinfo {author} {\bibfnamefont {H.}~\bibnamefont {Firouzjahi}},
  \ and\ \bibinfo {author} {\bibfnamefont {M.}~\bibnamefont {Sasaki}},\ }\href
  {\doibase 10.1209/0295-5075/101/39001} {\bibfield  {journal} {\bibinfo
  {journal} {Europhys. Lett.}\ }\textbf {\bibinfo {volume} {101}},\ \bibinfo
  {pages} {39001} (\bibinfo {year} {2013})},\ \Eprint
  {http://arxiv.org/abs/1210.3692} {arXiv:1210.3692 [astro-ph.CO]} \BibitemShut
  {NoStop}%
%%CITATION = ARXIV:1210.3692;%%
\bibitem [{\citenamefont {Martin}\ \emph {et~al.}(2013)\citenamefont {Martin},
  \citenamefont {Motohashi},\ and\ \citenamefont {Suyama}}]{Martin:2012pe}%
  \BibitemOpen
  \bibfield  {author} {\bibinfo {author} {\bibfnamefont {J.}~\bibnamefont
  {Martin}}, \bibinfo {author} {\bibfnamefont {H.}~\bibnamefont {Motohashi}}, \
  and\ \bibinfo {author} {\bibfnamefont {T.}~\bibnamefont {Suyama}},\ }\href
  {\doibase 10.1103/PhysRevD.87.023514} {\bibfield  {journal} {\bibinfo
  {journal} {Phys. Rev.}\ }\textbf {\bibinfo {volume} {D87}},\ \bibinfo {pages}
  {023514} (\bibinfo {year} {2013})},\ \Eprint {http://arxiv.org/abs/1211.0083}
  {arXiv:1211.0083 [astro-ph.CO]} \BibitemShut {NoStop}%
%%CITATION = ARXIV:1211.0083;%%
\bibitem [{\citenamefont {Motohashi}\ \emph {et~al.}(2015)\citenamefont
  {Motohashi}, \citenamefont {Starobinsky},\ and\ \citenamefont
  {Yokoyama}}]{Motohashi:2014ppa}%
  \BibitemOpen
  \bibfield  {author} {\bibinfo {author} {\bibfnamefont {H.}~\bibnamefont
  {Motohashi}}, \bibinfo {author} {\bibfnamefont {A.~A.}\ \bibnamefont
  {Starobinsky}}, \ and\ \bibinfo {author} {\bibfnamefont {J.}~\bibnamefont
  {Yokoyama}},\ }\href {\doibase 10.1088/1475-7516/2015/09/018} {\bibfield
  {journal} {\bibinfo  {journal} {JCAP}\ }\textbf {\bibinfo {volume} {1509}},\
  \bibinfo {pages} {018} (\bibinfo {year} {2015})},\ \Eprint
  {http://arxiv.org/abs/1411.5021} {arXiv:1411.5021 [astro-ph.CO]} \BibitemShut
  {NoStop}%
%%CITATION = ARXIV:1411.5021;%%
\bibitem [{\citenamefont {Motohashi}\ and\ \citenamefont
  {Starobinsky}(2017{\natexlab{a}})}]{Motohashi:2017aob}%
  \BibitemOpen
  \bibfield  {author} {\bibinfo {author} {\bibfnamefont {H.}~\bibnamefont
  {Motohashi}}\ and\ \bibinfo {author} {\bibfnamefont {A.~A.}\ \bibnamefont
  {Starobinsky}},\ }\href {\doibase 10.1209/0295-5075/117/39001} {\bibfield
  {journal} {\bibinfo  {journal} {Europhys. Lett.}\ }\textbf {\bibinfo {volume}
  {117}},\ \bibinfo {pages} {39001} (\bibinfo {year} {2017}{\natexlab{a}})},\
  \Eprint {http://arxiv.org/abs/1702.05847} {arXiv:1702.05847 [astro-ph.CO]}
  \BibitemShut {NoStop}%
%%CITATION = ARXIV:1702.05847;%%
\bibitem [{\citenamefont {Motohashi}\ and\ \citenamefont
  {Starobinsky}(2017{\natexlab{b}})}]{Motohashi:2017vdc}%
  \BibitemOpen
  \bibfield  {author} {\bibinfo {author} {\bibfnamefont {H.}~\bibnamefont
  {Motohashi}}\ and\ \bibinfo {author} {\bibfnamefont {A.~A.}\ \bibnamefont
  {Starobinsky}},\ }\href@noop {} {\  (\bibinfo {year} {2017}{\natexlab{b}})},\
  \Eprint {http://arxiv.org/abs/1704.08188} {arXiv:1704.08188 [astro-ph.CO]}
  \BibitemShut {NoStop}%
%%CITATION = ARXIV:1704.08188;%%
\bibitem [{\citenamefont {Kadota}\ \emph {et~al.}(2005)\citenamefont {Kadota},
  \citenamefont {Dodelson}, \citenamefont {Hu},\ and\ \citenamefont
  {Stewart}}]{Kadota:2005hv}%
  \BibitemOpen
  \bibfield  {author} {\bibinfo {author} {\bibfnamefont {K.}~\bibnamefont
  {Kadota}}, \bibinfo {author} {\bibfnamefont {S.}~\bibnamefont {Dodelson}},
  \bibinfo {author} {\bibfnamefont {W.}~\bibnamefont {Hu}}, \ and\ \bibinfo
  {author} {\bibfnamefont {E.~D.}\ \bibnamefont {Stewart}},\ }\href {\doibase
  10.1103/PhysRevD.72.023510} {\bibfield  {journal} {\bibinfo  {journal} {Phys.
  Rev.}\ }\textbf {\bibinfo {volume} {D72}},\ \bibinfo {pages} {023510}
  (\bibinfo {year} {2005})},\ \Eprint {http://arxiv.org/abs/astro-ph/0505158}
  {arXiv:astro-ph/0505158 [astro-ph]} \BibitemShut {NoStop}%
%%CITATION = ASTRO-PH/0505158;%%
\bibitem [{\citenamefont {Adshead}\ \emph {et~al.}(2012)\citenamefont
  {Adshead}, \citenamefont {Dvorkin}, \citenamefont {Hu},\ and\ \citenamefont
  {Lim}}]{Adshead:2011jq}%
  \BibitemOpen
  \bibfield  {author} {\bibinfo {author} {\bibfnamefont {P.}~\bibnamefont
  {Adshead}}, \bibinfo {author} {\bibfnamefont {C.}~\bibnamefont {Dvorkin}},
  \bibinfo {author} {\bibfnamefont {W.}~\bibnamefont {Hu}}, \ and\ \bibinfo
  {author} {\bibfnamefont {E.~A.}\ \bibnamefont {Lim}},\ }\href {\doibase
  10.1103/PhysRevD.85.023531} {\bibfield  {journal} {\bibinfo  {journal} {Phys.
  Rev.}\ }\textbf {\bibinfo {volume} {D85}},\ \bibinfo {pages} {023531}
  (\bibinfo {year} {2012})},\ \Eprint {http://arxiv.org/abs/1110.3050}
  {arXiv:1110.3050 [astro-ph.CO]} \BibitemShut {NoStop}%
%%CITATION = ARXIV:1110.3050;%%
\bibitem [{\citenamefont {Miranda}\ \emph {et~al.}(2012)\citenamefont
  {Miranda}, \citenamefont {Hu},\ and\ \citenamefont
  {Adshead}}]{Miranda:2012rm}%
  \BibitemOpen
  \bibfield  {author} {\bibinfo {author} {\bibfnamefont {V.}~\bibnamefont
  {Miranda}}, \bibinfo {author} {\bibfnamefont {W.}~\bibnamefont {Hu}}, \ and\
  \bibinfo {author} {\bibfnamefont {P.}~\bibnamefont {Adshead}},\ }\href
  {\doibase 10.1103/PhysRevD.86.063529} {\bibfield  {journal} {\bibinfo
  {journal} {Phys. Rev.}\ }\textbf {\bibinfo {volume} {D86}},\ \bibinfo {pages}
  {063529} (\bibinfo {year} {2012})},\ \Eprint {http://arxiv.org/abs/1207.2186}
  {arXiv:1207.2186 [astro-ph.CO]} \BibitemShut {NoStop}%
%%CITATION = ARXIV:1207.2186;%%
\bibitem [{\citenamefont {Miranda}\ \emph {et~al.}(2016)\citenamefont
  {Miranda}, \citenamefont {Hu}, \citenamefont {He},\ and\ \citenamefont
  {Motohashi}}]{Miranda:2015cea}%
  \BibitemOpen
  \bibfield  {author} {\bibinfo {author} {\bibfnamefont {V.}~\bibnamefont
  {Miranda}}, \bibinfo {author} {\bibfnamefont {W.}~\bibnamefont {Hu}},
  \bibinfo {author} {\bibfnamefont {C.}~\bibnamefont {He}}, \ and\ \bibinfo
  {author} {\bibfnamefont {H.}~\bibnamefont {Motohashi}},\ }\href {\doibase
  10.1103/PhysRevD.93.023504} {\bibfield  {journal} {\bibinfo  {journal} {Phys.
  Rev.}\ }\textbf {\bibinfo {volume} {D93}},\ \bibinfo {pages} {023504}
  (\bibinfo {year} {2016})},\ \Eprint {http://arxiv.org/abs/1510.07580}
  {arXiv:1510.07580 [astro-ph.CO]} \BibitemShut {NoStop}%
%%CITATION = ARXIV:1510.07580;%%
\bibitem [{\citenamefont {Motohashi}\ and\ \citenamefont
  {Hu}(2015)}]{Motohashi:2015hpa}%
  \BibitemOpen
  \bibfield  {author} {\bibinfo {author} {\bibfnamefont {H.}~\bibnamefont
  {Motohashi}}\ and\ \bibinfo {author} {\bibfnamefont {W.}~\bibnamefont {Hu}},\
  }\href {\doibase 10.1103/PhysRevD.92.043501} {\bibfield  {journal} {\bibinfo
  {journal} {Phys. Rev.}\ }\textbf {\bibinfo {volume} {D92}},\ \bibinfo {pages}
  {043501} (\bibinfo {year} {2015})},\ \Eprint
  {http://arxiv.org/abs/1503.04810} {arXiv:1503.04810 [astro-ph.CO]}
  \BibitemShut {NoStop}%
%%CITATION = ARXIV:1503.04810;%%
\bibitem [{\citenamefont {Bellini}\ and\ \citenamefont
  {Sawicki}(2014)}]{Bellini:2014fua}%
  \BibitemOpen
  \bibfield  {author} {\bibinfo {author} {\bibfnamefont {E.}~\bibnamefont
  {Bellini}}\ and\ \bibinfo {author} {\bibfnamefont {I.}~\bibnamefont
  {Sawicki}},\ }\href {\doibase 10.1088/1475-7516/2014/07/050} {\bibfield
  {journal} {\bibinfo  {journal} {JCAP}\ }\textbf {\bibinfo {volume} {1407}},\
  \bibinfo {pages} {050} (\bibinfo {year} {2014})},\ \Eprint
  {http://arxiv.org/abs/1404.3713} {arXiv:1404.3713 [astro-ph.CO]} \BibitemShut
  {NoStop}%
%%CITATION = ARXIV:1404.3713;%%
\bibitem [{\citenamefont {Hu}\ and\ \citenamefont {Joyce}(2017)}]{Hu:2016wfa}%
  \BibitemOpen
  \bibfield  {author} {\bibinfo {author} {\bibfnamefont {W.}~\bibnamefont
  {Hu}}\ and\ \bibinfo {author} {\bibfnamefont {A.}~\bibnamefont {Joyce}},\
  }\href {\doibase 10.1103/PhysRevD.95.043529} {\bibfield  {journal} {\bibinfo
  {journal} {Phys. Rev.}\ }\textbf {\bibinfo {volume} {D95}},\ \bibinfo {pages}
  {043529} (\bibinfo {year} {2017})},\ \Eprint
  {http://arxiv.org/abs/1612.02454} {arXiv:1612.02454 [astro-ph.CO]}
  \BibitemShut {NoStop}%
%%CITATION = ARXIV:1612.02454;%%
\bibitem [{\citenamefont {Chen}\ \emph
  {et~al.}(2013{\natexlab{a}})\citenamefont {Chen}, \citenamefont {Firouzjahi},
  \citenamefont {Namjoo},\ and\ \citenamefont {Sasaki}}]{Chen:2013aj}%
  \BibitemOpen
  \bibfield  {author} {\bibinfo {author} {\bibfnamefont {X.}~\bibnamefont
  {Chen}}, \bibinfo {author} {\bibfnamefont {H.}~\bibnamefont {Firouzjahi}},
  \bibinfo {author} {\bibfnamefont {M.~H.}\ \bibnamefont {Namjoo}}, \ and\
  \bibinfo {author} {\bibfnamefont {M.}~\bibnamefont {Sasaki}},\ }\href
  {\doibase 10.1209/0295-5075/102/59001} {\bibfield  {journal} {\bibinfo
  {journal} {Europhys. Lett.}\ }\textbf {\bibinfo {volume} {102}},\ \bibinfo
  {pages} {59001} (\bibinfo {year} {2013}{\natexlab{a}})},\ \Eprint
  {http://arxiv.org/abs/1301.5699} {arXiv:1301.5699 [hep-th]} \BibitemShut
  {NoStop}%
%%CITATION = ARXIV:1301.5699;%%
\bibitem [{\citenamefont {Huang}\ and\ \citenamefont
  {Wang}(2013)}]{Huang:2013lda}%
  \BibitemOpen
  \bibfield  {author} {\bibinfo {author} {\bibfnamefont {Q.-G.}\ \bibnamefont
  {Huang}}\ and\ \bibinfo {author} {\bibfnamefont {Y.}~\bibnamefont {Wang}},\
  }\href {\doibase 10.1088/1475-7516/2013/06/035} {\bibfield  {journal}
  {\bibinfo  {journal} {JCAP}\ }\textbf {\bibinfo {volume} {1306}},\ \bibinfo
  {pages} {035} (\bibinfo {year} {2013})},\ \Eprint
  {http://arxiv.org/abs/1303.4526} {arXiv:1303.4526 [hep-th]} \BibitemShut
  {NoStop}%
%%CITATION = ARXIV:1303.4526;%%
\bibitem [{\citenamefont {Chen}\ \emph
  {et~al.}(2013{\natexlab{b}})\citenamefont {Chen}, \citenamefont {Firouzjahi},
  \citenamefont {Komatsu}, \citenamefont {Namjoo},\ and\ \citenamefont
  {Sasaki}}]{Chen:2013eea}%
  \BibitemOpen
  \bibfield  {author} {\bibinfo {author} {\bibfnamefont {X.}~\bibnamefont
  {Chen}}, \bibinfo {author} {\bibfnamefont {H.}~\bibnamefont {Firouzjahi}},
  \bibinfo {author} {\bibfnamefont {E.}~\bibnamefont {Komatsu}}, \bibinfo
  {author} {\bibfnamefont {M.~H.}\ \bibnamefont {Namjoo}}, \ and\ \bibinfo
  {author} {\bibfnamefont {M.}~\bibnamefont {Sasaki}},\ }\href {\doibase
  10.1088/1475-7516/2013/12/039} {\bibfield  {journal} {\bibinfo  {journal}
  {JCAP}\ }\textbf {\bibinfo {volume} {1312}},\ \bibinfo {pages} {039}
  (\bibinfo {year} {2013}{\natexlab{b}})},\ \Eprint
  {http://arxiv.org/abs/1308.5341} {arXiv:1308.5341 [astro-ph.CO]} \BibitemShut
  {NoStop}%
%%CITATION = ARXIV:1308.5341;%%
\bibitem [{\citenamefont {Hirano}\ \emph {et~al.}(2016)\citenamefont {Hirano},
  \citenamefont {Kobayashi},\ and\ \citenamefont {Yokoyama}}]{Hirano:2016gmv}%
  \BibitemOpen
  \bibfield  {author} {\bibinfo {author} {\bibfnamefont {S.}~\bibnamefont
  {Hirano}}, \bibinfo {author} {\bibfnamefont {T.}~\bibnamefont {Kobayashi}}, \
  and\ \bibinfo {author} {\bibfnamefont {S.}~\bibnamefont {Yokoyama}},\ }\href
  {\doibase 10.1103/PhysRevD.94.103515} {\bibfield  {journal} {\bibinfo
  {journal} {Phys. Rev.}\ }\textbf {\bibinfo {volume} {D94}},\ \bibinfo {pages}
  {103515} (\bibinfo {year} {2016})},\ \Eprint
  {http://arxiv.org/abs/1604.00141} {arXiv:1604.00141 [astro-ph.CO]}
  \BibitemShut {NoStop}%
%%CITATION = ARXIV:1604.00141;%%
\end{thebibliography}%

\end{document}